\nonstopmode

\newcommand{\re}{{\mathrm{Re}\,}}
\newcommand{\im}{{\mathrm{Im}\,}}
\newcommand{\eV}{\U{eV}}
\newcommand{\mul}{\cdot}
\newcommand{\bohr}{\, a_0}
\newcommand{\Hartree}{\U{E_{\mathit h}}}
\newcommand{\angstrom}{\U{\hbox{\AA}}}
\newcommand{\invcm}{\U{cm^{-1}}}
\newcommand{\ie}{i.e.{}}
\newcommand{\eg}{e.g.{}}
\newcommand{\etal}{\textit{et al.}}
\newcommand{\U}[1]{\,{\rm{#1}}}
\newcommand{\I}[1]{_{\mathrm{#1}}}
\newcommand{\imag}{{\rm i}}
\newcommand{\euler}{\mathrm e}
\newcommand{\mat}[1]{\hbox{\boldmath{$#1$}\unboldmath}}
\newcommand{\Sum}{\sum\limits}
\newcommand{\Int}{\int\limits}
\newcommand{\transpose}{{}^{\textrm{\scriptsize T}}}
\newcommand{\unitop}{\hat{\mathbbm{1}}}
\newcommand{\differential}{\>\mathrm d}
\newcommand{\bra}[1]{\left<\right.\!#1\!\left.\right|}
\newcommand{\ket}[1]{\left|\right.\!#1\!\left.\right>}
\newcommand{\bracket}[2]{\left<\right.\!#1\left.\right|#2\!\left.\right>}
\newcommand{\diag}{\textbf{diag}}
\newcommand{\CFtBr}{CF$_3$Br}
\newcommand{\expectval}[1]{\left<#1\right>}
\newcommand{\cleb}[2]{C(#1 ; #2)}
\newcommand{\Tr}{{\mathrm{Tr}}\,}
\newcommand{\E}[1]{\times 10^{#1}}
\newcommand{\dfrac}[2]{{\displaystyle{#1}\over\displaystyle{#2}}}

\documentclass[pra,twocolumn,showpacs,nopreprintnumbers]{revtex4}
\usepackage{bm,bbm,amssymb,subeqnarray,graphicx}
\usepackage[nativepdf,bookmarks,bookmarksopen,bookmarksnumbered,raiselinks,%
pdftitle={Theory of x-ray absorption by laser-aligned symmetric-top molecules},%
pdfauthor={Christian Buth, Robin Santra},%
pdfsubject={Chemical physics},%
pdfkeywords={ab initio, laser alignment, x rays, photoabsorption,
cross section, polarization dependence, perturbation theory, nonperturbative, %
laser-matter interaction, Br2, bromine molecule, density matrix, %
symmetric top, rigid rotor}]{hyperref}

\begin{document}
\title{Theory of x-ray absorption by laser-aligned symmetric-top molecules}
\author{Christian Buth}
\author{Robin Santra}
\affiliation{Argonne National Laboratory, Argonne, Illinois~60439, USA}
\date{January 10, 2007}

\begin{abstract}
We devise a theory of x-ray absorption by symmetric-top molecules which
are aligned by an intense optical laser.
Initially, the density matrix of the system is composed of the
electronic ground state of the molecules and a thermal ensemble
of rigid-rotor eigenstates.
We formulate equations of motion of the two-color (laser plus
x~rays) rotational-electronic problem.
The interaction with the laser is assumed to be nonresonant;
it is described by an electric dipole polarizability tensor.
X-ray absorption is approximated as a one-photon process.
It is shown that the equations can be separated such that the interaction
with the laser can be treated independently of the x~rays.
The laser-only density matrix is propagated numerically.
After each time step, the x-ray absorption is calculated.
We apply our theory
to study adiabatic alignment of bromine molecules~(Br$_2$).
The required dynamic polarizabilities are determined using the \emph{ab initio}
linear response methods
coupled-cluster singles~(CCS),
second-order approximate coupled-cluster singles and doubles~(CC2),
and coupled-cluster singles and doubles~(CCSD).
For the description of x-ray absorption on the
$\sigma\I{g} \, 1s \to \sigma\I{u} \, 4p$~resonance,
a parameter-free two-level model is used for the electronic structure
of the molecules.
Our theory opens up novel perspectives for the quantum control of
x-ray radiation.
\end{abstract}

%
%
%

\maketitle

\section{Introduction}

We study the interaction of an intense optical laser and hard
x~rays with molecules, a so-called two-color problem.
The laser is assumed to be sufficiently intense to influence molecular
rotation, however, without vibrationally and electronically exciting
the molecules in a gas sample.
Specifically, we regard x-ray absorption by laser-aligned symmetric-top
molecules.
On the one hand, this leads to dynamical information of molecular rotation;
on the other hand, this opens up novel ways to control x~rays.

Such a combination of an optical laser with x~rays has proved very beneficial
already in various setups.
For intense laser pulses, ionization of atoms takes place, producing
aligned, unoccupied atomic orbitals.
They are probed by using XUV~light or x~rays to excite an inner-shell
electron into them~\cite{Santra:SO-06,Loh:QS-07,Santra:SF-07}.
For slightly lower laser intensities, one only modifies the atomic Rydberg
manifold leading to electromagnetically induced transparency
for x~rays~\cite{Buth:TX-07,Buth:ET-07,Santra:SF-07}.

Laser-alignment of molecules was discovered experimentally
by Normand~\etal~\cite{Normand:LI-92} and was later analyzed
theoretically by Friedrich and Herschbach~\cite{Friedrich:AT-95}.
Other alignment techniques, however, did exist before, \eg, with
strong static electric and magnetic fields for molecules with
permanent electric and magnetic dipole moments,
respectively~\cite{Stapelfeldt:AM-03,Seideman:NA-05}.
Friedrich and Herschbach~\cite{Friedrich:AT-95} considered alignment with
a continuous-wave laser which lead to eigenstates
of the coupled problem of molecular rotation in the laser field.
Ortigoso~\etal{}~\cite{Ortigoso:TE-99} carried out a systematic theoretical
and computational study of laser-aligned linear molecules.
They allowed for pulsed lasers and rotational wave packets to
investigate the transition from a thermal ensemble
to the continuous-wave solution of Friedrich and
Herschbach~\cite{Friedrich:AT-95}.
Aside of this so-called adiabatic alignment, there is also transient
(or impulsive) alignment~\cite{Stapelfeldt:AM-03,Seideman:NA-05} which means that
the molecules are ``kicked'' with an intense, short laser pulse and
thereafter evolve freely.
This leads to brief, recurring periods when all molecules in a
sample are aligned~\cite{Stapelfeldt:AM-03,Seideman:NA-05}.
The work of Ortigoso~\etal~\cite{Ortigoso:TE-99} was extended recently
to symmetric-top molecules by Hamilton~\etal~\cite{Hamilton:AS-05}.

The alignment of a gaseous sample causes the individual
molecules to basically rotate in the same way.
Hence, at sufficiently low temperature, the x-ray absorption signal
reflects the properties of a single molecule rather than a
rotational average.
Adiabatic laser-alignment of molecules is the only choice when
the molecules are to be studied with x~rays from third-generation
synchrotron sources due to long x-ray pulses of
about~$100 \U{ps}$~\cite{Thompson:XR-01} which are
several orders of magnitude longer than the time scale of transient
alignment of about~$1 \U{ps}$~\cite{Stapelfeldt:AM-03,Seideman:NA-05}.
Particularly suited for the study of alignment is a dependence of the
absorption cross section on the angle between the linear x-ray
polarization and the figure axis of symmetric-top molecules.
Specifically, such a linear dichroism can be found on single, isolated
resonances in the x-ray absorption near-edge fine
structure~(XANES) region~\cite{Rehr:TA-00,Als-Nielsen:EM-01}.
For instance, above the carbon $K$~edge, there is a
$\sigma^*$~shape resonance in hydrocarbons.
The dependence of the cross section on the resonance on the orientation of the
molecules was theoretically and experimentally studied with x~rays
for molecules adsorbed on surfaces~\cite{Haack:SR-00}.

Using x~rays to quantify molecular alignment has several advantages over
conventional methods.
Typically, molecular alignment is detected by, first, cleaving a
molecular bond with one laser
pulse and ionizing the fragments with a very intense other laser
pulse~\cite{Stapelfeldt:AM-03}.
The latter pulse leads to highly ionized fragments which undergo
Coulomb explosion.
Due to the complicated nature of the interaction of a laser
with a molecule, this is a very intricate technique to study molecular
alignment;
\eg, recently, the strong-field ionization probability of aligned
molecules was measured and a nontrivial, molecule-specific angular
dependence was found~\cite{Pavicic:DM-07}.
Therefore, this technique is prone to systematic errors and
it is highly desirable to have an alternative
method for the detection of molecular alignment.

In order to theoretically investigate the two-color problem,
it is crucial to analyze and exploit the three
different time scales of the involved physical processes which are
assumed to be well-separated by a few orders of magnitude from
each other.
First, the x-ray oscillations define the fastest timescale.
For example, at the $K$~edge of a bromine atom, $\omega\I{X} =
13474 \eV$~\cite{Thompson:XR-01},
a cycle period amounts to~$\frac{2\pi}{\omega\I{X}} = 0.30694 \U{as}$.
Second, x-ray absorption leads to core-hole formation.
A $K$~vacancy in a bromine atom relaxes with
a decay time of~$\frac{1}{\Gamma} = 0.264 \U{fs}$, \ie, the
line width is~$\Gamma = 2.49 \eV$~\cite{Campbell:WA-01}.
This relaxation defines the coherence time scale, \ie, the time scale on which
x-ray amplitudes instead of intensities must be considered
and thus interference effects in x-ray absorption may occur
due to molecular rotation.
Third, the time scale of molecular rotation, and x-ray and laser
envelopes;
the quickest of them defines the third scale.
This last scale varies with the choice of parameters;
situations may arise, \eg, high laser intensity, where the separation
of the third time scale from the second is no longer sufficient and a joint
treatment of the processes on both time scales is required.
For this work, we expect the third time scale to be of a few tens of
femtoseconds to several picoseconds, \ie, a separate treatment of
the physical processes on the second and on third time scales is feasible.
Because of this distinction of three time scales, the slower processes can
be taken to be constant while approximations to the faster processes
are obtained.
Using the approximations to the faster process, the variation
on the slower time scale can be determined.

In this work, we study the influence of molecular alignment on the
absorption cross section for linearly polarized x~rays.
We devise a density matrix formalism for rotational motion of
symmetric-top molecules under the influence of a laser.
The laser-only equations are then extended to comprise the interaction
with the x~rays.
Our work lays the theoretical foundation for our computational studies
of bromotrifluoromethane~(\CFtBr{}) in Ref.~\onlinecite{Santra:SF-07}
and the joint investigation of~\CFtBr{} with experimentalists in
Ref.~\onlinecite{Peterson:XR-up}.

The paper is structured as follows.
Section~\ref{sec:theory} discusses the theoretical foundation.
We start with the laser-only problem in Sec.~\ref{sec:laseralign};
the full two-color problem (laser plus x~rays) is examined
in Sec.~\ref{sec:xray_probe}.
The x-ray absorption cross section and derived quantities are discussed
in Sec.~\ref{sec:photoabsorb}.
In Sec.~\ref{sec:elstrmod}, a two-level electronic structure model is
developed for bromine molecules.
Computational details are given in Sec.~\ref{sec:compdet} and the
results are presented in Sec.~\ref{sec:results}.
Conclusions are drawn in Sec.~\ref{sec:conclusion}.

Our equations are formulated in atomic units~\cite{Szabo:MQC-89},
where $1 \U{hartree} = 1 \Hartree$ is the unit of energy,
$1 \, t_0$~is the unit of time,
and $1 \U{bohr} = 1 \bohr$ is the unit of length.
The Boltzmann constant~$k\I{B}$ is unity and
$1 \Hartree = 3.15775 \E{5} \U{K}$~is the unit of temperature.
Intensities are given in units of~$1 \Hartree \> t_0^{-1} \, a_0^{-2}
= 6.43641 \times 10^{15} \U{W \, cm^{-2}}$.

\section{Theory}
\label{sec:theory}
\subsection{Laser alignment of symmetric-top molecules}
\label{sec:laseralign}
\subsubsection{The freely rotating molecule}

We assume the Born-Oppenheimer approximation and the rigid-rotor approximation
which implies that rotational and electronic degrees of freedom are
decoupled~\cite{Kroto:MR-75,Szabo:MQC-89} and molecular vibrations
are excluded~\cite{Kroto:MR-75}.
Without interactions that reestablish a coupling,
electronic structure and molecular rotation can be treated independently.
We deal in the following with two reference frames:
one is the laboratory or space-fixed frame where the coordinates are denoted
by~$x, y, z$.
The other frame is the molecule-fixed or body-fixed frame with the
coordinates~$a, b, c$~\cite{Kroto:MR-75}.

The molecular rotation is described by the rigid-rotor Hamiltonian which
reads~\cite{Kroto:MR-75}
\begin{equation}
  \label{eq:molrot}
  \hat H\I{r} = A \hat J_A^2 + B \hat J_B^2 + C \hat J_C^2 \; ,
\end{equation}
where $A$, $B$, and~$C$ are rotational constants.
For a symmetric-top molecule, $\hat H\I{r}$ commutes with the square of the
angular momentum operator~$\hat{\vec J^2}$, the $z$~component of
the angular momentum in the space-fixed frame~$\hat J_z$,
and the $c$~component of the angular momentum in the body-fixed
frame~$\hat J_c$~\cite{Rose:ET-57,Kroto:MR-75,Zare:AM-88}.
The three operators give rise to the rotational quantum numbers~$J$,
$K$, and~$M$, respectively, with~$K, M = -J, \ldots, J$ for integer~$J \geq 0$.
We assume throughout a prolate symmetric top, $A > B = C$, where the
axis~$A$ is identified with the figure axis~$c$.
For an oblate symmetric top, $A = B > C$ and the axis~$C$ becomes
the figure axis~$c$~\cite{Kroto:MR-75}.
In this case, all~$A$ need to be replaced by~$C$ in what follows.
We obtain from Eq.~(\ref{eq:molrot}) the Hamiltonian
\begin{equation}
  \label{eq:symolrot}
  \hat H\I{r} = B \hat{\vec J^2} + (A-B) \hat J_c^2
\end{equation}
with the eigenvalues~\cite{Kroto:MR-75,Zare:AM-88}
\begin{equation}
  \label{eq:syrotev}
  E_{JK} = B \, J(J+1) + (A-B) \, K^2 \; .
\end{equation}
The eigenfunctions of the time-independent Schr\"odinger equation
with the Hamiltonian~(\ref{eq:symolrot}) are the symmetric-rotor
functions~\cite{Zare:AM-88}
\begin{equation}
  \label{eq:syrofunc}
  \psi_{JMK}(\varphi, \vartheta, \chi) = (-1)^{M-K} \sqrt{\frac{2J+1}
    {8\pi^2}} \, D_{-M,-K}^J(\varphi, \vartheta, \chi) \; .
\end{equation}
Here, $\varphi$, $\vartheta$, $\chi$ denote the three Euler angles
that connect the space- and body-fixed reference
frames~\cite{Kroto:MR-75,Zare:AM-88} and $D^J_{-M,-K}(\varphi,
\vartheta, \chi)$ represents elements of the Wigner matrix for
angular momentum~$J$~\cite{Rose:ET-57,Kroto:MR-75,Zare:AM-88}.
The eigenfunctions of the general asymmetric rotor can be expanded
in this basis~\cite{Stapelfeldt:AM-03,Seideman:NA-05}.

The molecules in a gas sample are initially not in a definite quantum state
but are described by a thermodynamic ensemble~\cite{Kroto:MR-75,Reichl:SP-04}.
Therefore, a density operator formalism is
beneficial~\cite{Blum:DM-96,Reichl:SP-04}.
Let~$T$ be the rotational equilibrium temperature of the ensemble
and~$\beta = \frac{1}{T}$.
Then, the density operator of a freely rotating symmetric-top
molecule reads in its stationary eigenbasis~(\ref{eq:syrofunc})
\begin{equation}
  \label{eq:thdenseig}
  \hat\varrho\I{nucl} = \Sum_{J, K, M} \ket{J K M} \, w_{JK}
    \, \bra{J K M} \; ,
\end{equation}
with the density matrix elements
\begin{equation}
  \label{eq:dmfree}
  w_{JK} = g_I(J,K) \> \frac{\euler^{-\beta E_{JK}}}{Z} \; .
\end{equation}
The weights~$g_I(J, K)$ reflect the nuclear spin
statistics~\cite{Townes:MS-55,Kroto:MR-75} and
\begin{equation}
  \label{eq:partfunc}
  Z = \Sum_{J} (2J+1) \Sum_{K=-J}^J g_I(J, K) \, \euler^{-\beta E_{JK}}
\end{equation}
is the partition function.

\subsubsection{Interaction with a nonresonant laser}

Let the laser radiation of angular frequency~$\omega\I{L}$ be linearly
polarized along the $z$~axis, \ie, the polarization vector
is~$\vec e\I{L} = \vec e_z$.
The laser electric field is
\begin{equation}
  \label{eq:laserelfield}
  \vec E\I{L}(t) = \varepsilon\I{L}(t) \, \vec e\I{L}  \,
    \cos(\omega\I{L} \, t) \; .
\end{equation}
Here, $\varepsilon\I{L}(t) = \sqrt{8 \pi \alpha I\I{L}(t)}$ is the
envelope amplitude, $\alpha$ is the fine-structure constant, and
$I\I{L}(t)$ is the cycle-averaged intensity of the laser pulse at time~$t$.
We assume a Gaussian envelope with peak intensity~$I\I{L,0}$
and a [full width at half maximum~(FWHM)] duration of~$\tau\I{L}$
\begin{equation}
  \label{eq:GaussEnv}
  I\I{L}(t) = I\I{L,0} \> \euler^{-4 \ln 2 (\frac{t}{\tau\I{L}})^2} \; .
\end{equation}
The peak intensity of the pulse (at~$t = 0$) depends on the laser pulse
energy~$E\I{pulse}$ as~$I\I{L,0} = 2 \, \sqrt{\frac{\ln 2}{\pi}} \>
\frac{E\I{pulse}}{\tau\I{L}} \, X(0)$.
The factor~$X(0) = \frac{4\ln 2}{\pi \varrho\I{L}^2}$
represents the peak of a Gaussian radial profile of a FWHM width
of~$\varrho\I{L}$.
The treatment of~$X(\varrho)$, with $\varrho$ being the radius perpendicular
to the beam axis, would involve a number of computations with decreasing
intensity.

The interaction of molecules with an optical laser of a frequency far
away from electronic transitions is described using the
dynamic dipole polarizability of the molecules~\cite{Kroto:MR-75,%
Craig:MQ-84,Bishop:MV-90,Merzbacher:QM-98};
it is given in terms of a symmetric, rank two, Cartesian
tensor~\cite{Goodbody:CT-82} which we denote
by~$\mat\alpha^{\mathrm S}(\omega\I{L})$ in the space-fixed frame.
The interaction Hamiltonian reads~\cite{Merzbacher:QM-98}
\begin{equation}
  \label{eq:intHamLas}
  \hat H\I{L}(t) = -\frac{1}{2} \vec E\I{L}(t)^{\dagger}
    \mat\alpha^{\mathrm S}(\omega\I{L}) \vec E\I{L}(t) \; .
\end{equation}
After temporal averaging of Eq.~(\ref{eq:intHamLas}) over an
optical cycle, we obtain~$\hat H\I{L}(t)$ for the case~$\vec e\I{L}
= \vec e\I{z}$ as
\begin{equation}
  \label{eq:lasIntHam}
  \hat H\I{L}(t) = -2\pi\alpha \, I\I{L}(t) \, \alpha^{\mathrm
    S}_{zz}(\omega\I{L}) \; .
\end{equation}
The dipole polarizability tensor is decomposed~\cite{Rose:ET-57,%
Sobelman:AS-79,Zare:AM-88} into a linear
combination of a rank zero spherical tensor and the five components of a
rank two spherical tensor~\cite{footnote1};
the $zz$~component is rewritten as~\cite{Sobelman:AS-79}
\begin{equation}
  \label{eq:polsph}
  \alpha^{\mathrm S}_{zz}(\omega\I{L}) = \alpha^{\mathrm S\prime}_{0,0}
    (\omega\I{L}) + \alpha^{\mathrm S\prime}_{2,0}(\omega\I{L}) \; ,
\end{equation}
where we use a prime accent to indicate spherical tensor components.
The term $\alpha^{\mathrm S\prime}_{0,0}(\omega\I{L})
= \frac{1}{3} \Tr \mat \alpha^{\mathrm S}(\omega\I{L})$ is invariant
under rotations and thus does not have any impact on molecular alignment.

We would like to express~$\alpha^{\mathrm S}_{zz}(\omega\I{L})$ in terms of
the dynamic polarizability tensor in the molecule-fixed
frame~$\mat\alpha^{\mathrm M}(\omega\I{L})$ because the dynamic polarizability
tensor is time independent in this frame.
The transformation from the space-fixed to the body-fixed reference
frame reads for spherical tensor operators~\cite{Rose:ET-57,Zare:AM-88}
\begin{equation}
  \label{eq:coordrot}
  \alpha^{\mathrm S\prime}_{2,m}(\omega\I{L}) = \Sum_{m'=-2}^2
    D^{2\,*}_{mm'}(\varphi, \vartheta, \chi) \> \alpha^{\mathrm
    M\prime}_{2,m'}(\omega\I{L}) \; .
\end{equation}
We choose the axes of the molecule-fixed coordinate system to be the
principal axes of~$\mat\alpha^{\mathrm M}(\omega\I{L})$.
Then, we have for a symmetric-top rotor, $\mat\alpha^{\mathrm M}(\omega\I{L})
= \diag[\alpha_{\perp}(\omega\I{L}), \alpha_{\perp}(\omega\I{L}),
\alpha_{\parallel}(\omega\I{L})]$.
There are only two distinct diagonal elements:
$\alpha_{\parallel}(\omega\I{L})$ is the polarizability parallel to the
figure axis of the molecule and $\alpha_{\perp}(\omega\I{L})$~is the
polarizability perpendicular to it.
The components of the spherical tensors in Eq.~(\ref{eq:polsph}) of
rank zero and rank two~\cite{Sobelman:AS-79}
are found from the relations~$\alpha^{\mathrm M\prime}_{0,0}(\omega\I{L})
= \frac{1}{3} [2\alpha_{\perp}(\omega\I{L}) + \alpha_{\parallel}(\omega\I{L})]$,
$\alpha^{\mathrm M\prime}_{2,0}(\omega\I{L})
= \frac{2}{3} [\alpha_{\parallel}(\omega\I{L}) -
\alpha_{\perp}(\omega\I{L})]$, and
$\alpha^{\mathrm M\prime}_{2,\pm 1}(\omega\I{L})
= \alpha^{\mathrm M\prime}_{2,\pm 2}(\omega\I{L}) = 0$.
The following formulas are expressed more compactly in terms of the
average dynamical polarizability
\begin{equation}
  \label{eq:polave}
  \bar\alpha(\omega\I{L}) = \frac{1}{3} \Sum_{i=a,b,c} \alpha^{\mathrm
    M}_{ii}(\omega\I{L}) = \frac{2}{3} \, \alpha_{\perp}(\omega\I{L})
    + \frac{1}{3} \, \alpha_{\parallel}(\omega\I{L})
\end{equation}
and the dynamical polarizability anisotropy
\begin{equation}
  \label{eq:polani}
  \Delta \alpha(\omega\I{L}) = \alpha_{\parallel}(\omega\I{L})
    - \alpha_{\perp}(\omega\I{L}) \; .
\end{equation}

Using Eqs.~(\ref{eq:lasIntHam}), (\ref{eq:polsph}), (\ref{eq:coordrot}),
(\ref{eq:polave}), (\ref{eq:polani}), and~\cite{Hamilton:AS-05}
\begin{equation}
  \label{eq:3D-rule}
  \begin{array}{rcl}
    \bra{JKM} D^{J''}_{m'm} \ket{J'K'M'} = \sqrt{\dfrac{2J+1}{2J'+1}} && \\
      \times \cleb{JJ''J'}{KmK'} \, \cleb{JJ''J'}{Mm'M'} \; , &&
  \end{array}
\end{equation}
the matrix elements of the interaction Hamiltonian with respect to the
symmetric-rotor functions~(\ref{eq:syrofunc})
are found to be~\cite{Rose:ET-57,Zare:AM-88}
\begin{equation}
  \label{eq:lasermat}
  \begin{array}{rcl}
    &&{}\displaystyle \bra{JKM} \hat H\I{L}(t) \ket{J'K'M'} \\
    &=& \displaystyle - \frac{4}{3} \pi \alpha \, I\I{L}(t) \, \Delta
      \alpha(\omega\I{L}) \, \delta_{KK'} \, \delta_{MM'} \\
    &&{} \displaystyle \times \sqrt{\frac{2J+1}{2J'+1}} \, \cleb{J2J'}{K0K}
      \, \cleb{J2J'}{M0M} \\
    &&{} \displaystyle - 2\pi \alpha \, I\I{L}(t) \, \bar
      \alpha(\omega\I{L}) \, \delta_{JJ'} \, \delta_{KK'} \, \delta_{MM'} \; .
  \end{array}
\end{equation}
Here, $\cleb{j_1 j_2 j_3}{m_1 m_2 m_3} = \bracket{j_1 m_1, j_2 m_2}{j_3 m_3}$
denote Clebsch-Gordan coefficients~\cite{Rose:ET-57,Zare:AM-88}.
The last summand in Eq.~(\ref{eq:lasermat}) causes only a global
energy shift;
it has no impact on the rotational dynamics and may thus be dropped.
Nota bene, the interaction matrix of a symmetric-top molecule for a
linearly polarized laser is diagonal in~$K$ and $M$.

\subsubsection{Equation of motion for the rotational dynamics}
\label{sec:EOMlaser}

The time-evolution operator in the interaction picture---indicated by the
subscript~I---from~$t\I{s}$ to~$t$ is denoted
by~$\hat U\I{nucl, I}(t, t\I{s})$~\cite{footnote2}.
The relation of a Schr\"odinger-picture state vector~$\ket{\psi,t}$
and operator~$\hat O(t)$ to interaction picture state
vector~$\ket{\psi,t}\I{I}$ and operator~$\hat O\I{I}(t)$
is~\cite{Merzbacher:QM-98}
\begin{subeqnarray}
  \label{eq:opSchrInt}
  \slabel{eq:waveSchrInt}
  \ket{\psi,t}\I{I} &=& \euler^{\imag \hat H_0 t} \, \ket{\psi,t} \; , \\
  \slabel{eq:operatorSchrInt}
  \hat O\I{I}(t) &=& \euler^{\imag \hat H_0 t} \, \hat O(t) \,
    \euler^{-\imag \hat H_0 t} \; ,
\end{subeqnarray}
with the Hamiltonian of the noninteracting system~$\hat H_0$.
Here, $\hat H_0 = \hat H\I{r}$;
later, from Sec.~\ref{sec:xray_probe} onwards, $\hat H_0$~will also comprise
the nuclear repulsion plus electronic Hamiltonian~\cite{Szabo:MQC-89}
of the molecule~$\hat H\I{M}$, \eg, in Eq.~(\ref{eq:totHam}).

At $t = t\I{s}$, there is no laser and the density operator of the
ensemble of symmetric-top molecules in the interaction picture
results from Eqs.~(\ref{eq:thdenseig}) and (\ref{eq:operatorSchrInt}).
For later times, the density operator is formally given by~\cite{Blum:DM-96}
\begin{equation}
  \label{eq:timenucldens}
  \hat\varrho\I{nucl, I}(t) = \hat U\I{nucl, I}(t, t\I{s}) \,
    \hat\varrho\I{nucl, I}(t\I{s}) \, \hat U\I{nucl,
    I}^{\dagger}(t, t\I{s}) \; .
\end{equation}
The time evolution can be recast as follows
\begin{equation}
  \label{eq:timepsi}
  \ket{\psi_{JKM}, t}\I{I} = \hat U\I{nucl, I}(t, t\I{s}) \ket{JKM} \; .
\end{equation}
Let
\begin{equation}
  \label{eq:wfcoeff}
  c_{J'J}^{(KM)}(t) = \bracket{J'KM}{\psi_{JKM}, t}\I{I}
\end{equation}
be the time-dependent expansion coefficients.
As the matrix representation~(\ref{eq:lasermat}) of the interaction
Hamiltonian with the laser is diagonal with respect to~$K$ and $M$,
so is the coefficient matrix~(\ref{eq:wfcoeff}).
With Eqs.~(\ref{eq:thdenseig}), (\ref {eq:dmfree}), (\ref{eq:timenucldens}),
(\ref{eq:timepsi}), and (\ref{eq:wfcoeff}),
the density operator of a thermodynamic ensemble of symmetric-top
molecules in a laser field has the following structure in the
interaction picture:
\begin{equation}
  \label{eq:densnuclprop}
  \begin{array}{rcl}
    \displaystyle \hat\varrho\I{nucl,I}(t) &=& \displaystyle \Sum_{J, J',
      J'', K, M} \ket{J'KM} \, c_{J'J}^{(KM)}(t) \> w_{JK} \\
    &&{} \displaystyle \times c_{J''J}^{(KM) \, *}(t) \, \bra{J''KM} \; .
  \end{array}
\end{equation}
Due to Eqs.~(\ref{eq:syrotev}) and (\ref{eq:operatorSchrInt}), the density
operator in the Schr\"odinger picture simply results by multiplying
the right-hand side of Eq.~(\ref{eq:densnuclprop}) with the picture
transformation factor~$\euler^{-\imag \, \omega_{J'J''} \, t}$ where
[Eq.~(\ref{eq:syrotev})]
\begin{equation}
  \label{eq:VSchrInt}
  \omega_{J'J''} = E_{J'K} - E_{J''K} = B [J'(J'+1) - J''(J''+1)] \; ,
\end{equation}
for all~$K$, \ie, the density matrix elements are
\begin{equation}
  \label{eq:densnuclpropschroe}
  \begin{array}{rcl}
    \displaystyle \varrho_{J' J''}^{(KM)}(t) &=& \displaystyle \bra{J' K M}
      \hat\varrho\I{nucl}(t) \ket{J'' K M} \\
    &=& \displaystyle \euler^{-\imag \, \omega_{J'J''} \, t} \bra{J' K M}
      \hat\varrho\I{nucl,I}(t) \ket{J'' K M} \; .
  \end{array}
\end{equation}
The equation of motion of an ensemble of molecules in a laser field
is given by the Liouville equation~\cite{Blum:DM-96,Reichl:SP-04}
\begin{equation}
  \label{eq:Liouville}
  \imag \, \dot{\hat \varrho}\I{nucl, I}(t) = [ \hat H\I{L, I}(t),
    \hat \varrho\I{nucl, I}(t) ] \; .
\end{equation}
Using Eqs.~(\ref{eq:opSchrInt}) and (\ref{eq:densnuclprop}), one reduces
Eq.~(\ref{eq:Liouville}) to equations of motion for the
coefficients~(\ref{eq:wfcoeff})
\begin{equation}
  \label{eq:coeffEOM}
  \begin{array}{rcl}
    \imag \, \dot c_{J'J}^{(K M)}(t) &=& \Sum_{J''} \bra{J' K M}
      \hat H\I{L}(t) \ket{J'' K M} \\
    &&{} \times \euler^{\imag \, \omega_{J'J''} \, t} c_{J''J}^{(K M)}(t) \; .
  \end{array}
\end{equation}
These equations form a system of coupled, first-order ordinary
differential equations~\cite{Golub:SC-92}.
In order to solve Eq.~(\ref{eq:coeffEOM}), the time
integral~$\Int^t_{t\I{s}} \dot c_{J'J}^{(KM)}(t') \differential t'$
needs to be found.
By comparing Eq.~(\ref{eq:thdenseig}) with Eq.~(\ref{eq:densnuclprop}),
we obtain the initial conditions~$c_{J'J}^{(KM)}(t\I{s}) = \delta_{J'J}$
with the help of Eqs.~(\ref{eq:dmfree}), (\ref{eq:timepsi}),
and (\ref{eq:wfcoeff}).
Employing standard numerical methods for systems of differential
equations~\cite{Golub:SC-92}, an approximate solution can be
determined.
In doing so, the full propagation interval is decomposed into~$N\I{t}$
subintervals which are sufficiently short such that the
coefficients~$c_{J''J}^{(KM)}(t)$ on the right-hand side of
Eq.~(\ref{eq:coeffEOM}) and the laser envelope, Eq.~(\ref{eq:GaussEnv}), can
be assumed to be constant on each subinterval.

To quantify the degree of alignment along the laser polarization axis,
one typically uses the expectation value~\cite{Blum:DM-96,%
Stapelfeldt:AM-03,Reichl:SP-04,Seideman:NA-05}
\begin{equation}
  \label{eq:cos2theta}
  \langle \, \cos^2 \vartheta \, \rangle(t)
    = \Tr [\hat\varrho\I{nucl}(t) \, \cos^2 \vartheta] \; .
\end{equation}
Here, $\vartheta$ refers to the Euler angle that determines the tilt
between the $c$~axis and the $z$~axis of the two coordinate
systems~\cite{Kroto:MR-75,Zare:AM-88}.
Using Eq.~(\ref{eq:3D-rule}), the necessary matrix elements
of~$\cos^2 \vartheta = \frac{2}{3} \, D_{00}^2(\varphi, \vartheta, \chi)
+ \frac{1}{3}$ in the basis of symmetric-rotor functions~(\ref{eq:syrofunc})
read
\begin{equation}
  \begin{array}{rl}
    \bra{JKM} \cos^2 \vartheta \ket{J'K'M'}
      = \delta_{KK'} \, \delta_{MM'} \bigl[
      \dfrac{1}{3} \, \delta_{JJ'} & \\
    {} + \dfrac{2}{3} \, \sqrt{\dfrac{2J+1}{2J'+1}} \>
      \cleb{J2J'}{K0K} \, \cleb{J2J'}{M0M} \bigr] \; . &
  \end{array}
\end{equation}

\subsection{X-ray probe}
\label{sec:xray_probe}
\subsubsection{Interaction with x~rays}

The x-ray pulse has the same functional form as the laser
pulse~(\ref{eq:laserelfield})---however with~L replaced by~X---with
the same, Gaussian, form for the envelope~$\varepsilon\I{X}(t)$
as the laser pulse~(\ref{eq:GaussEnv}).
The semiclassical Hamiltonian for the interaction of a molecule
with x~rays of angular frequency~$\omega\I{X}$
reads~\cite{Meystre:QO-91}
\begin{equation}
  \label{eq:xrayint}
  \hat H\I{X}(t) = - (\hat{\vec d} \cdot \vec e\I{X,M}) \, \varepsilon\I{X}(t)
    \, \cos (\omega\I{X} \, t) = \hat h\I{X}(t) \, \cos (\omega\I{X} \, t) \; .
\end{equation}
Here, $\hat{\vec d}$~is the electric dipole operator.
The x~rays are assumed to be polarized linearly along the
direction~$\vec e\I{X,M}$.
Both vectors are given with respect to the molecule-fixed, Cartesian
coordinate system.

For simplicity, we transform the vectors from the Cartesian to the spherical
basis~\cite{Rose:ET-57,Zare:AM-88}.
This change of coordinate system is indicated by a prime accent.
It leaves the scalar product formally the same with the sum over Cartesian
vector components~$j \in \{x, y, z\}$ and $\alpha \in \{a, b, c\}$ replaced
by a sum over projection quantum numbers~$m \in \{-1, 0, 1\}$.
Using a coordinate system transformation~\cite{Rose:ET-57,%
Zare:AM-88} analogous to Eq.~(\ref{eq:coordrot})
\begin{equation}
  \label{eq:xpolbody}
  (\vec e^{\>\prime}\I{X,M})_m = \Sum_{m'=-1}^1 D^{1}_{m'm}(\varphi,
    \vartheta, \chi) (\vec e^{\>\prime}\I{X})_{m'} \; ,
\end{equation}
we express the body-fixed x-ray polarization vector~$\vec e^{\>\prime}\I{X,M}$
in terms of the space-fixed x-ray polarization vector~$\vec e^{\>\prime}\I{X}$.

The dependence of~$\hat H\I{X}(t)$ in Eq.~(\ref{eq:xrayint}) on
electronic and rotational degrees of freedom constitutes a coupling
of both.
Thus, the interaction with the x~rays needs to be evaluated in terms
of the direct product basis~\cite{Cohen:QM-77}
\begin{equation}
  \label{eq:dirprod}
  \ket{JKM, i} = \ket{JKM} \otimes \ket{i} \; ,
\end{equation}
consisting of the symmetric-rotor states~$\ket{JKM}$ and the electronic
states~$\ket{i}$.
Here, $i = 0$ denotes the electronic ground state.
We obtain~\cite{footnote5}
\begin{equation}
  \label{eq:H_X_matel}
  \begin{array}{rcl}
    &&\bra{JKM, i} \hat H\I{X}(t) \ket{J'K'M', i'} \\
    &=& -(\vec d^{\>\prime\,*}_{ii'} \mul \vec s^{\>\prime}_{JKM,J'K'M'}) \>
    \varepsilon\I{X}(t) \, \cos (\omega\I{X} \, t) \; .
  \end{array}
\end{equation}
The electric dipole matrix elements in the body-fixed
reference frame are designated by
\begin{equation}
  \label{eq:diptransme}
  \vec d^{\>\prime}_{ii'} = \bra{i} \hat{\vec d^{\>\prime}} \ket{i'} \; ;
\end{equation}
the nuclear rotation factor is defined by
\begin{equation}
  \label{eq:srotmatel}
  \vec s^{\>\prime}_{JKM,J'K'M'} = \bra{JKM} \vec e^{\>\prime}\I{X,M}
    \ket{J'K'M'} \; .
\end{equation}
The matrix elements~(\ref{eq:srotmatel}) are easily evaluated using
Eqs.~(\ref{eq:syrofunc}), (\ref{eq:3D-rule}), and (\ref{eq:xpolbody}).

\subsubsection{The two-color problem}

The joint Hamiltonian for nuclear dynamics, electronic structure, and
the interaction with laser and x~rays reads
\begin{equation}
  \label{eq:totHam}
  \hat H = [\hat H\I{r} + \hat H\I{L}(t)] \otimes \unitop\I{el}
    + \unitop\I{nucl} \otimes \hat H\I{M} + \hat H\I{X}(t) \; .
\end{equation}
Here, $\unitop\I{el}$~is the unit operator in the space of electronic
states and $\unitop\I{nucl}$~is the unit operator in
the manifold of symmetric-rotor states~(\ref{eq:syrofunc}).
Further, $\hat H\I{r}$~is given in Eq.~(\ref{eq:symolrot}),
$\hat H\I{L}(t)$~in Eq.~(\ref{eq:lasIntHam}), and $\hat H\I{X}(t)$~in
Eq.~(\ref{eq:xrayint}).
Finally, $\hat H\I{M}$~denotes the fixed-nuclei molecular
Hamiltonian~\cite{Szabo:MQC-89}.

To describe the time-evolution of the interaction of a molecule
with a laser and x~rays, we extend the density operator formalism of
Sec.~\ref{sec:EOMlaser} to include electronic degrees of freedom.
Initially, the molecule is in the electronic ground state~$\ket{0}$.
The corresponding stationary density operator simply reads~\cite{Blum:DM-96}
\begin{equation}
  \label{eq:densopel}
  \hat \varrho\I{el} = \ket{0}\bra{0} \; .
\end{equation}
The initial rotational density operator is given by a thermal
distribution~(\ref{eq:thdenseig}).
Then, the joint rotational-electronic problem is initially described by the
direct product~\cite{Cohen:QM-77,Blum:DM-96}
\begin{equation}
  \hat \varrho = \hat \varrho\I{nucl} \otimes \hat \varrho\I{el}
\end{equation}
because, by assumption, for~$t \leq t\I{s}$, we have~$\hat H\I{X}(t) = 0$
in Eq.~(\ref{eq:totHam}).
The relevant equations from
Sec.~\ref{sec:laseralign}---Eqs.~(\ref{eq:timenucldens}),
(\ref{eq:timepsi}), (\ref{eq:wfcoeff}), (\ref{eq:VSchrInt}), (\ref{eq:Liouville}),
and (\ref{eq:coeffEOM})---can
now be generalized to comprise also the electronic degree of freedom.
To this end, we replace the nuclear rotational states~$\ket{JKM}$
with the direct product states~$\ket{JKM,i}$ from
Eq.~(\ref{eq:dirprod})~\cite{Cohen:QM-77}.
Formally, this replacement means only that another index, namely, $i$~needs
to be added to the existing equations.
The only thing one has to pay attention to is the fact that the interaction with
the x~rays---unlike the interaction with the laser~(\ref{eq:lasermat})---is
generally not diagonal in~$K$ and $M$.
Hence, the coefficients~(\ref{eq:wfcoeff}) have to be replaced by
\begin{equation}
  \label{eq:wfxlcoeff}
  c_{JKM,i; J'K'M',0}(t) = \bracket{JKM,i}{\psi_{J'K'M',0}, t}\I{I} \; .
\end{equation}
By assumption, the molecule is initially in the electronic ground
state~($i=0$).
The energies of rotational-electronic states are
\begin{equation}
  \label{eq:roelen}
  E_{JK,i} = E_{JK} + E_i - \imag \frac{\Gamma}{2} \, (1 - \delta_{0 i}) \; ,
\end{equation}
with~$E_{JK}$ denoting the energy of nuclear rotational
states~(\ref{eq:syrotev}) and $E_i$~being the
energy of electronic states of~$\hat H\I{M}$ in Eq.~(\ref{eq:totHam}).
We consider x-ray absorption to produce only electronically excited states
($i \neq 0$) with a single core hole which is an excellent approximation for
the moderate x-ray intensities at third-generation synchrotron sources.
The core vacancies relax by x-ray fluorescence and Auger
decay~\cite{Thompson:XR-01,Als-Nielsen:EM-01}.
The associated decay width is indicated as~$\Gamma$;
it is assumed not to vary for different core-excited states.
This is an excellent approximation when the core vacancy is always in
the same shell and has always the same angular momentum.

We obtain the equation of motion for the joint problem in
analogy to Eq.~(\ref{eq:coeffEOM})
\begin{equation}
  \label{eq:elCoeffEOM}
  \begin{array}{rcl}
    &&\displaystyle  \imag \, \dot c_{J'K'M',i'; JKM,0}(t) \\
    &=&\displaystyle  \Sum_{J'',K'',M'',i''} \bra{J'K'M', i'}
    \hat H\I{L}(t) \otimes \unitop\I{el} \\
    &&\displaystyle{} + \hat H\I{X}(t) \ket{J'' K'' M'', i''}
      \euler^{\imag \, \omega_{J'K',i'; J''K'',i''} \, t} \\
    &&\displaystyle{} \times
      c_{J''K''M'',i''; JKM,0}(t) \; .
  \end{array}
\end{equation}
In contrast to Eq.~(\ref{eq:coeffEOM}), here the sum extends
also over~$K''$, $M''$, and~$i''$.
The picture-transformation factor [compare with Eq.~(\ref{eq:VSchrInt})]
becomes using Eq.~(\ref{eq:roelen}):
\begin{equation}
  \label{eq:roelomega}
  \begin{array}{rcl}
   &&\displaystyle \omega_{J'K',i'; J''K'',i''} = E_{J'K',i'}
     - E_{J''K'',i''} \\
   &=&\displaystyle E_{J'K'} - E_{J''K''} + E_{i'} - E_{i''}
     - \imag \frac{\Gamma}{2} \, (\delta_{0 i''} - \delta_{0 i'}) \; .
  \end{array}
\end{equation}

\subsubsection{Recovering the laser-only equations}
\label{sec:reclasonly}

Let us consider first the case~$i' = 0$ in Eq.~(\ref{eq:elCoeffEOM}).
From Eqs.~(\ref{eq:lasermat}) and (\ref{eq:dirprod}) follows
\begin{equation}
  \label{eq:H_L_dirprod}
  \begin{array}{rcl}
    \bra{JKM,i} \hat H\I{L}(t) \ket{J'K'M',i'} = \delta_{KK'} \, \delta_{MM'}
      \, \delta_{ii'} && \\
    {} \times \bra{JKM} \hat H\I{L}(t) \ket{J'KM} \; . &&
  \end{array}
\end{equation}
We arrive at
\begin{equation}
  \label{eq:noexcite}
  \begin{array}{rcl}
    &&\displaystyle \imag \, \dot c_{J'K'M',0; JKM,0}(t) \\
    &=&\displaystyle \Sum_{J''} \bra{J'K'M'} \hat H\I{L}(t)
      \ket{J''K'M'} \\
    &&\displaystyle{} \times \euler^{\imag \, \omega_{J'J''} \, t}
      \> c_{J''K'M',0; JKM,0}(t) \\
    &&\displaystyle{} + \Sum_{J'',K'',M''} \bra{J'K'M', 0} \hat H\I{X}(t)
      \ket{J''K''M'', 0} \\
    &&\displaystyle{} \times \euler^{\imag \, \omega_{J'K',0; J''K'',0} \, t}
      \> c_{J''K''M'',0; JKM,0}(t) \\
    &&\displaystyle{} + \Sum_{\scriptstyle J'',K'',M'',i'' \atop \scriptstyle
      i'' \neq 0} \bra{J'K'M', 0} \hat H\I{X}(t) \ket{J''K''M'', i''} \\
    &&\displaystyle{} \times \euler^{\imag \, \omega_{J'K',0; J''K'',i''} \, t}
      \> c_{J''K''M'',i''; JKM,0}(t) \; .
  \end{array}
\end{equation}
The first term on the left-hand side of Eq.~(\ref{eq:noexcite})
constitutes the interaction of the laser with the rotational degrees of
freedom.
It is the same as in Eq.~(\ref{eq:coeffEOM}) in the expanded direct
product space~(\ref{eq:dirprod}).
Similarly, the second term in Eq.~(\ref{eq:noexcite}) represents transitions
between rotational states due to the x~rays.
The third term in Eq.~(\ref{eq:noexcite}) describes an x-ray induced
recombination of an excited electron with the core hole.

The time evolution of this expanded system of differential
equations~(\ref{eq:noexcite}) is treated in analogy to the laser-only
equations in Sec.~\ref{sec:EOMlaser} by approximating~$\Int^t_{t\I{s}}
\dot c_{J'K'M',0; JKM,0}(t') \differential t'$.
For this purpose, we decompose~$[t\I{s}, t]$ into $N\I{t}$ subintervals.
We require, in addition to the laser-only
case~(\ref{eq:coeffEOM}), that the envelope of
the x-ray pulse~$\hat h\I{X}(t')$ on the right-hand side of
Eq.~(\ref{eq:xrayint}) is approximately constant over the subintervals.
(Please refer to the introduction concerning a discussion of the involved
physical time scales which need to be taken into account.)
For our forthcoming approximations to be appropriate,
the duration of the subintervals are also required
to be much longer than core-hole lifetimes~$\frac{1}{\Gamma}$ and the
duration of x-ray oscillations~$\frac{2\pi}{\omega\I{X}}$

For molecules without permanent electric dipole moment, \ie, $\vec
d^{\>\prime}_{00} = 0$ [Eq.~(\ref{eq:diptransme})], the second term in
Eq.~(\ref{eq:noexcite}) vanishes exactly because the matrix
element~(\ref{eq:H_X_matel}) is zero.
Otherwise, this term leads to a negligible integral on the time
subintervals due to rapid oscillations, Eq.~(\ref{eq:H_X_matel}),
of~$\euler^{\imag \, \omega_{J'K',0; J''K'',0} \, t} \cos(\omega\I{X} \, t)$.
Thereby, we assume moderate x-ray intensities, \ie, the total x-ray
absorption probability for the full propagation interval is tiny.
This also means that the third term is tiny because it depends on
coefficients with~$i \neq 0$.
Omitting the third term in Eq.~(\ref{eq:noexcite}) implies
that two- (and multi)-photon processes for x~rays are neglected.

With these arguments, the second and third terms in Eq.~(\ref{eq:noexcite})
are omitted leading to the laser-only equations of motion~(\ref{eq:coeffEOM}).
In other words, we treat the interaction with the laser independently of the
x~rays in what follows.

\subsubsection{Electronic excitation due to x-ray absorption}

Let us now consider the case~$i' \neq 0$ in Eq.~(\ref{eq:elCoeffEOM}).
With Eq.~(\ref{eq:H_L_dirprod}), we find
\begin{equation}
  \label{eq:xrayexcite}
  \begin{array}{rcl}
    &&\displaystyle \imag \, \dot c_{J'K'M',i'; JKM,0}(t) \\
    &=&\displaystyle \Sum_{J'',K'',M''} \bra{J'K'M', i'} \hat H\I{X}(t)
      \ket{J''K''M'', 0} \\
    &&\displaystyle{} \times \euler^{\imag \, \omega_{J'K',i'; J''K'',0} \, t}
      \> c_{J''K''M'',0; JKM,0}(t) \\
    &&\displaystyle{} + \Sum_{J''} \bra{J'K'M'} \hat H\I{L}(t) \ket{J''K'M'} \\
    &&\displaystyle{} \times \euler^{\imag \, \omega_{J'J''} \, t}
      \> c_{J''K'M',i'; JKM,0}(t) \\
    &&\displaystyle{} + \Sum_{\scriptstyle J'',K'',M'',i'' \atop\scriptstyle i''
      \neq 0} \bra{J'K'M', i'} \hat H\I{X}(t) \ket{J''K''M'', i''} \\
    &&\displaystyle{} \times \euler^{\imag \, \omega_{J'K',i'; J''K'',i''} \, t}
      \> c_{J''K''M'',i''; JKM,0}(t) \; .
  \end{array}
\end{equation}
The meaning of the individual summands on the right-hand side of the equation
is as follows.
The first term represents x-ray absorption by a ground-state electron
leading to electronic excitation of the molecule.
The second term quantifies laser-induced transitions between rotational
sublevels of an electronically excited state.
Finally, the third term describes transitions by x-ray absorption between
rotational-electronic core-hole states.
Since we neglect x-ray nonlinear processes, we set this term to zero.
One expects the second term in Eq.~(\ref{eq:xrayexcite}) to
make a vanishingly small contribution, too, which is shown in
the ensuing paragraphs.

We need an approximate expression for~$c_{J''K'M',i'; JKM,0}(t)$ in
the second summand on the right-hand side of Eq.~(\ref{eq:xrayexcite}).
We carry out the time integration of the right-hand side---thereby neglecting the
third term---in analogy to the previous Sec.~\ref{sec:reclasonly}
by decomposing the interval~$[t\I{s},t]$ into $N\I{t}$ subintervals,
letting~$t = t_{N\I{t}}$:
\begin{equation}
  \label{eq:approxexcoeff}
  \begin{array}{rcl}
    &&\displaystyle c_{J'K'M',i'; JKM,0}(t) \\
    &\approx&\displaystyle \frac{1}{\imag} \Sum_{J'',K'',M''}
      \Sum_{n=0}^{N\I{t}-1}
      \bra{J'K'M', i'} \hat h\I{X}(t_n) \ket{J''K''M'', 0} \\
    &&\displaystyle{}\qquad \times c_{J''K''M'',0; JKM,0}(t_n) \>
      I^{(1)}_{J'K',i'; J''K'',0}(t_n,t_{n+1}) \\
    &&\displaystyle{} + \frac{1}{\imag} \Sum_{J''} \Sum_{n=0}^{N\I{t}-1}
      \bra{J'K'M'} \hat H\I{L}(t_n) \ket{J''K'M'} \\
    &&\displaystyle{}\qquad \times  c_{J''K'M',i'; JKM,0}(t_n) \>
      I^{(2)}_{J'J''}(t_n,t_{n+1}) \; .
  \end{array}
\end{equation}
The remaining time integrals~$I^{(1)}_{J'K',i'; J''K'',0}(t_n,t_{n+1})$
and $I^{(2)}_{J'J''}(t_n,t_{n+1})$ in Eq.~(\ref{eq:approxexcoeff}) can be
carried out immediately;
rapidly oscillating contributions are discarded employing
the rotating-wave approximation~(RWA)~\cite{Meystre:QO-91}.
We find
\begin{equation}
  \label{eq:fastint}
  \begin{array}{cl}
    &\displaystyle I^{(1)}_{J'K',i'; J''K'',0}(t_n,t_{n+1}) \\
    \displaystyle =&\displaystyle \Int_{t_n}^{t_{n+1}} \cos
      (\omega\I{X} \, t') \> \euler^{\imag \, \omega_{J'K',i'; J''K'',0} \, t'}
      \differential t' \\
    \displaystyle \approx &\displaystyle \frac{\euler^{\imag
      (\omega_{J'K',i'; J''K'',0} - \omega\I{X}) \, t_{n+1}}
      - \euler^{\imag \, (\omega_{J'K',i'; J''K'',0} - \omega\I{X}) \, t_n}}
      {2 \imag \, (\omega_{J'K',i'; J''K'',0} - \omega\I{X})} \\
    \displaystyle \approx &\displaystyle \frac{\euler^{\imag \,
      (\omega_{J'K',i'; J''K'',0} - \omega\I{X}) \, t_{n+1}}}
      {2 \imag \, (\omega_{J'K',i'; J''K'',0} - \omega\I{X})}
  \end{array}
\end{equation}
and
\begin{equation}
  \label{eq:slowint}
  \begin{array}{cl}
    &\displaystyle I^{(2)}_{J'J''}(t_n,t_{n+1})
     = \Int_{t_n}^{t_{n+1}} \euler^{\imag \,
      \omega_{J'J''} \, t'} \differential t' \\
    \displaystyle \approx & \displaystyle \euler^{\imag \, \omega_{J'J''} \,
      t_n} \> (t_{n+1} - t_n) \; .
  \end{array}
\end{equation}
In the last expression~(\ref{eq:slowint}), the
approximation~$\omega_{J'J''} \> (t_{n+1} - t_n) \ll 1 $ is
used, \ie, the rotational time scale is much longer than the duration
of the small intervals.

The coefficients~(\ref{eq:wfxlcoeff}) are defined with respect to
the interaction picture~(\ref{eq:opSchrInt}).
Yet the density operator eventually will be used in the Schr\"odinger
picture [see Eq.~(\ref{eq:eldensmat})], \ie, the coefficients involved are
\begin{equation}
  C_{J'K'M',i'; JKM,0}(t) = \euler^{-\imag \, E_{J'K',i'} \, t} \,
    c_{J'K'M',i'; JKM,0}(t) \; ,
\end{equation}
observing Eqs.~(\ref{eq:roelen}) and (\ref{eq:roelomega}).
The impact of this picture transformation factor on
Eq.~(\ref{eq:approxexcoeff}) is exclusively given by its impact on the
integrals~$I^{(1)}_{J'K',i'; J''K'',0}(t_n,t_{n+1})$ and
$I^{(2)}_{J'J''}(t_n,t_{n+1})$.
The first integral~(\ref{eq:fastint}) becomes with~$t = t_{N\I{t}}$:
\begin{equation}
  \label{eq:fastintsch}
  \begin{array}{cl}
    & \displaystyle \euler^{-\imag \, E_{J'K',i'} \, t} \,
      I^{(1)}_{J'K',i'; J''K'',0}(t_n,t_{n+1}) \\
    \displaystyle \approx & \displaystyle \delta_{N\I{t},n+1} \;
      \frac{\euler^{-\imag \, (E_{J'' K'', 0} + \omega\I{X}) \, t}}
           {2 \, \imag \, (\omega_{J'K',i'; J''K'',0} - \omega\I{X})} \; .
  \end{array}
\end{equation}

Due to~$\euler^{-\imag \, E_{J'K',i'} \, t} \, I^{(2)}_{J'J''}(t_n,t_{n+1})
\approx 0$, the second term in Eq.~(\ref{eq:approxexcoeff}) vanishes in the
Schr\"odinger picture because, by assumption, core-excited molecules
decay rapidly on a subinterval~[$\Gamma \> (t_{n+1} - t_n) \gg 1$]
and the laser-induced rotational response is negligible during the
decay time of~$\frac{1}{\Gamma}$.

Thus, Eq.~(\ref{eq:approxexcoeff}) goes over into
\begin{equation}
  \label{eq:approxexcoeffcoupe}
  \begin{array}{cl}
    &\displaystyle c_{J'K'M',i'; JKM,0}(t) \\
    \displaystyle \approx& \displaystyle \Sum_{J'',K'',M''}
      \bra{J'K'M', i'} \hat h\I{X}(t) \ket{J''K''M'', 0} \\
    & \displaystyle {}\times c_{J''K''M'',0; JKM,0}(t) \;
      \frac{\euler^{\imag \, (\omega_{J'K',i'; J''K'',0} - \omega\I{X}) \, t}}
           {-2 \, (\omega_{J'K',i'; J''K'',0} - \omega\I{X})} \; ,
  \end{array}
\end{equation}
which is the final form for the coefficients describing electronic excitations.
We arrive at the corresponding approximation of Eq.~(\ref{eq:xrayexcite}):
\begin{equation}
  \label{eq:xrayexcitecoupe}
  \begin{array}{cl}
    &\displaystyle \dot c_{J'K'M',i'; JKM,0}(t) \\
    \displaystyle = &\displaystyle  \frac{1}{\imag} \Sum_{J'',K'',M''}
      \bra{J'K'M', i'} \hat H\I{X}(t) \ket{J''K''M'', 0} \\
    &\displaystyle{} \times \euler^{\imag \, \omega_{J'K',i'; J''K'',0} \, t} \>
      c_{J''K''M'',0; JKM,0}(t) \; ,
  \end{array}
\end{equation}
which becomes identical to the time derivative of
Eq.~(\ref{eq:approxexcoeffcoupe}) after applying the RWA.

In analogy to Eqs.~(\ref{eq:densnuclprop}) and (\ref{eq:densnuclpropschroe}),
the Schr\"odinger-picture density matrix of the joint problem of
molecular rotation and electronic excitation reads
\begin{equation}
  \label{eq:eldensmat}
  \begin{array}{cl}
    &\displaystyle \varrho_{JKM, i; J'K'M', i'}(t) \\
    \displaystyle =& \displaystyle \Sum_{J'',K'',M''} w_{J''K''} \>
      C_{JKM,i;J''K''M'',0}(t) \\
    &\displaystyle{} \times C^*_{J'K'M',i';J''K''M'',0}(t) \; .
  \end{array}
\end{equation}
Explicit expressions for the density matrix elements
are determined from Eq.~(\ref{eq:eldensmat}) by inserting the
approximation~(\ref{eq:approxexcoeffcoupe}).
One arrives at density matrix elements between electronically excited
states~($i, i' \neq 0$)
\begin{equation}
  \label{eq:2coldm}
  \begin{array}{cl}
  &\displaystyle \varrho_{JKM, i; J'K'M', i'}(t) \\
  \displaystyle =& \displaystyle \frac{1}{4} \Sum_{J_1, J_2, K_1, M_1}
    \varrho_{J_1 J_2}^{(K_1 M_1)}(t) \>
    \frac{\bra{JKM , i } \hat h\I{X}(t) \ket{J_1 K_1 M_1, 0}}
         {\omega_{JK,i; J_1 K_1,0} - \omega\I{X}} \\
  &\displaystyle{} \times
    \frac{\bra{J'K'M', i'} \hat h\I{X}(t)\ket{J_2 K_1 M_1, 0}^*}
         {\omega^*_{J'K',i'; J_2 K_1,0} - \omega\I{X}} \; .
  \end{array}
\end{equation}
Observe that $\varrho_{J_1 J_2}^{(K_1 M_1)}(t)$ from
Eq.~(\ref{eq:densnuclpropschroe}) enters here which contains the
whole time evolution of the molecular rotation.
Hence, the laser-alignment problem and the interaction with the x~rays
can be treated independently.
The off-diagonal elements with either~$i = 0$ or $i' = 0$ of
Eq.~(\ref{eq:eldensmat})
are all negligible due to rapid oscillations.

\subsection{Photoabsorption cross section}
\label{sec:photoabsorb}

The probability of being in a core-excited state at time~$t$
follows from Eq.~(\ref{eq:eldensmat}) and is given by
\begin{equation}
  \label{eq:densprob}
  \begin{array}{rcl}
    Q(t) &=& \Sum_{\scriptstyle J,K,M \atop \scriptstyle i \neq 0}
      \varrho_{JKM, i; JKM, i}(t) \\
    &=& \Sum_{\scriptstyle J,K,M, \atop {\scriptstyle J'',K'',M'' \atop \scriptstyle
      i \neq 0}} w_{J''K''} \, |C_{JKM,i;J''K''M'',0}(t)|^2 \; ,
  \end{array}
\end{equation}
The rate of change of the probability of being in a core-excited
state then is
\begin{equation}
  \label{eq:expoprate}
  \dot Q(t) = - \Gamma \, Q(t) + \Gamma\I{abs}(t) \; ,
\end{equation}
in which
\begin{equation}
  \label{eq:xrayabsrate}
  \begin{array}{rcl}
    \Gamma\I{abs}(t) &=& 2 \Sum_{\scriptstyle J,K,M, \atop {\scriptstyle J'',K'',M''
      \atop \scriptstyle i \neq 0}} w_{J''K''} \  \re [ \; \euler^{- \Gamma t} \\
    &&{} \times \, \dot c^*_{JKM,i;J''K''M'',0}(t) \;
      c_{JKM,i;J''K''M'',0}(t)] \;.
  \end{array}
\end{equation}
To interpret Eq.~(\ref{eq:expoprate}), we assume, for the moment,
no decay of core-excited states, \ie, $\Gamma = 0$.
Then, the first term on the right-hand side vanishes and $\dot Q(t) =
\Gamma\I{abs}(t)$ holds.
Hence, $\dot Q(t)$ quantifies the rate of change of the excited
state population at time~$t$ due to the interaction with the x~rays.
In fact this means x-ray absorption because the nonlinear recombination
process of the excited electron with the core vacancy is neglected in
our equations [third term on the right-hand side of Eq.~(\ref{eq:xrayexcite})].
Consequently, $\Gamma\I{abs}(t)$ is the rate of x-ray absorption
at time~$t$.
With core-hole decay, \ie, $\Gamma \neq 0$, Eq.~(\ref{eq:expoprate}) has
the form of a rate equation~\cite{Meystre:QO-91}.
On the right-hand side of Eq.~(\ref{eq:expoprate}),
the first term denotes the rate of excited-state population loss due
to core-hole decay, whereas the second term represents the
rate of increase of excited-state population
due to x-ray absorption.

From the x-ray absorption rate~(\ref{eq:xrayabsrate}), the
photoabsorption cross section follows via
\begin{equation}
  \label{eq:xsectrate}
  \sigma(t) = \frac{\Gamma\I{abs}(t)}{J\I{X}(t)} \; .
\end{equation}
Here, $J_{\mathrm X}(t) = \frac{N\I{X}}{V\alpha}$ represents the flux
generated by $N\I{X}$~x-ray photons in the volume~$V$ that propagate
in the same direction with the speed of
light~$\frac{1}{\alpha}$~\cite{Craig:MQ-84}.
The fine-structure constant is designated by~$\alpha$.
We insert Eq.~(\ref{eq:approxexcoeffcoupe}) and the complex conjugate of
Eq.~(\ref{eq:xrayexcitecoupe}) into Eq.~(\ref{eq:xrayabsrate})
and use Eqs.~(\ref{eq:densnuclpropschroe}) and (\ref{eq:xsectrate}) to
obtain the instantaneous x-ray absorption cross section of a molecule
in a laser field
\begin{equation}
  \label{eq:xsect}
  \begin{array}{rcl}
    \displaystyle \sigma(t) &=& \displaystyle
      4 \pi \, \alpha \, \omega\I{X} \  \im \biggl[ \Sum_{J,J',K,M} \varrho_{J J'}^{(KM)}(t)
      \Sum_{J'',K'',M'', i''} \\
    &&\displaystyle{} \hspace{-1em} \times
      \frac{(\vec d^{\>\prime\,*}_{0i''} \mul \vec s_{J'KM, J''K''M''}^{\>\prime})
            (\vec d^{\>\prime\,*}_{i''0} \mul \vec s_{J''K''M'',JKM}^{\>\prime})}
      {E_{J''K''} + E_{i''} - \imag \, \Gamma / 2 - E_{JK} - E_0 - \omega\I{X}}
      \biggr] \; ,
  \end{array}
\end{equation}
with $\varepsilon\I{X}(t) = \sqrt{8\pi \, \alpha \, \omega\I{X}
\, J\I{X}(t)}$ and $I\I{X}(t) = \omega\I{X} \, J\I{X}(t)$
[see Eq.~(\ref{eq:laserelfield})].
The same expression~(\ref{eq:xsect}) is obtained from the theory of
Ref.~\onlinecite{Santra:SO-06} by assuming that the laser-only density matrix
is constant over short time intervals.
Formally Eq.~(\ref{eq:xsect}) resembles the expression~(40) in
Ref.~\onlinecite{Buth:TX-07} for the cross section of laser-dressed atoms.

Given the instantaneous cross section~(\ref{eq:xsect}), we can compute
experimentally more accessible quantities.
The cross correlation of the cross section and the x-ray pulse is defined
by~\cite{footnote3}
\begin{equation}
  \label{eq:crosscorr}
  P_i(\tau) = \Int_{-\infty}^{\infty} \sigma_i(t) \, J\I{X}(t-\tau)
    \differential t \; , \qquad i = \parallel, \perp, \mathrm{th} \; .
\end{equation}
Here, $\parallel$ and $\perp$ denote parallel and perpendicular
laser and x-ray polarization vectors whereas th~refers to a thermal
ensemble, \ie, no laser radiation is present.
Further, $P_i(\tau)$ represents the total probability of x-ray absorption
for a time delay of~$\tau$ between laser and x-ray pulses.
We define following ratios of cross correlations~(\ref{eq:crosscorr})
for parallel and perpendicular polarizations and a thermal ensemble:
\begin{subeqnarray}
  \label{eq:crossratios}
  \slabel{eq:crossparperp}
  R\I{\parallel / \perp      }(\tau) &=& \frac{P_{\parallel}(\tau)}
    {P_{\perp      }(\tau)} \; , \\
  R\I{\parallel / \mathrm{th}}(\tau) &=& \frac{P_{\parallel}(\tau)}
    {P_{\mathrm{th}}(\tau)} \; , \\
  R\I{\perp     / \mathrm{th}}(\tau) &=& \frac{P_{\perp    }(\tau)}
    {P_{\mathrm{th}}(\tau)} \; .
\end{subeqnarray}
The ratio~(\ref{eq:crossparperp}) has been studied
experimentally~\cite{Peterson:XR-up}.

One can rewrite the cross section~(\ref{eq:xsect}) to clearly exhibit
its dependence on the angle between laser and x-ray
polarizations~$\vartheta\I{LX}$.
We assume that laser polarization~$\vec e\I{L}$ and x-ray polarization~$\vec e\I{X}$
are initially parallel and directed along the $z$~axis.
Keeping the laser polarization fixed~$\vec e\I{L} = \vec e\I{z}$, one
obtains the rotated x-ray polarization vector~\cite{Kroto:MR-75}
with a passive rotation by~$\vartheta\I{LX}$ around the $y$~axis as
\begin{equation}
  \vec e\I{X} = \sin \vartheta\I{LX} \, \vec e\I{x}
    + \cos \vartheta\I{LX} \, \vec e\I{z} \; .
\end{equation}
The vector~$\vec e\I{X}$ needs to be transformed to the spherical
basis~\cite{Rose:ET-57,Zare:AM-88} and then be subjected to
Eq.~(\ref{eq:xpolbody}) to obtain the x-ray polarization vector in the
molecular frame~$\vec e^{\>\prime}\I{X, M}$.
Inserting this vector into Eq.~(\ref{eq:srotmatel}),
the angular dependence of the cross section~(\ref{eq:xsect}) reads
\begin{equation}
  \label{eq:angxsectdep}
  \begin{array}{rl}
    &\displaystyle{} \phantom{\sin^2 \vartheta\I{LX} \,}
      (\vec d^{\>\prime\,*}_{0i''} \mul \vec s_{J'KM, J''K''M''}^{\>\prime})
      (\vec d^{\>\prime\,*}_{i''0} \mul \vec s_{J''K''M'',  JKM}^{\>\prime}) \\
    \displaystyle =& \displaystyle{} \! \sin^2 \vartheta\I{LX} \,
      (\vec d^{\>\prime\,*}_{0i''} \mul \vec s_{J'KM, J''K''M''}^{\>\prime,\perp})
      (\vec d^{\>\prime\,*}_{i''0} \mul \vec s_{J''K''M'',  JKM}^{\>\prime,\perp}) \\
    + &\displaystyle{} \! \cos^2 \vartheta\I{LX} \,
      (\vec d^{\>\prime\,*}_{0i''} \mul \vec s_{J'KM, J''K''M''}^{\>\prime,\parallel})
      (\vec d^{\>\prime\,*}_{i''0} \mul \vec s_{J''K''M'',  JKM}^{\>\prime,\parallel}) \\
    + &\displaystyle{} \! \sin \vartheta\I{LX} \, \cos \vartheta\I{LX} \\
    & \displaystyle{} \phantom{\sin^2 \vartheta\I{LX} \!\!\!\!\!\!\!\!}
      \times \bigl[
      (\vec d^{\>\prime\,*}_{0i''} \mul \vec s_{J'KM, J''K''M''}^{\>\prime,\perp    })
      (\vec d^{\>\prime\,*}_{i''0} \mul \vec s_{J''K''M'',  JKM}^{\>\prime,\parallel}) \\
    &\displaystyle{}  \phantom{\sin^2 \vartheta\I{LX} \!\!\!\!\!} +
      (\vec d^{\>\prime\,*}_{0i''} \mul \vec s_{J'KM, J''K''M''}^{\>\prime,\parallel})
      (\vec d^{\>\prime\,*}_{i''0} \mul \vec s_{J''K''M'',  JKM}^{\>\prime,\perp    }) \bigr]
  \end{array}
\end{equation}
with the assignments~$(\vec s_{J'KM, J''K''M''}^{\>\prime, \perp})_m
= \frac{1}{\sqrt{2}} \bra{J'KM} -D_{1 \, m}^1 + D_{-1 \, m}^1\ket{J''K''M''}$
and $(\vec s_{J'KM, J''K''M''}^{\>\prime, \parallel})_m =
\bra{J'KM} D_{0 \, m}^1 \ket{J''K''M''}$ for~$m \in \{-1, 0, 1\}$.
The angular dependence of the cross section is expressed compactly by inserting
Eq.~(\ref{eq:angxsectdep}) into Eq.~(\ref{eq:xsect}):
\begin{equation}
  \label{eq:angxsect}
  \begin{array}{rcl}
    \displaystyle{} \sigma(\vartheta\I{LX}, t) &=& \displaystyle{} \sin^2
      \vartheta\I{LX} \, \sigma_{\perp}(t) + \cos^2 \vartheta\I{LX} \,
      \sigma_{\parallel}(t) \\
    &&\displaystyle{} + \sin \vartheta\I{LX} \, \cos \vartheta\I{LX} \,
      \sigma_{\perp, \parallel}(t) \; ,
  \end{array}
\end{equation}
where $\sigma_{\perp}(t)$, $\sigma_{\parallel}(t)$, and
$\sigma_{\perp, \parallel}(t)$ are the cross sections for parallel x-ray and
laser polarization vectors, perpendicular polarizations, and a cross
term.
Generally, $\sigma_{\perp, \parallel}(t)$ does not vanish.
However, we will show that there are no cross terms for the special case
of the two level model in the ensuing Sec.~\ref{sec:elstrmod}.

The angular dependence of~$\sigma(\vartheta\I{LX}, t)$ can be investigated
experimentally~\cite{Santra:SF-07,Peterson:XR-up} in terms of a wave
plate scan, \ie, the ratio
\begin{equation}
  \label{eq:waveplatefull}
  \begin{array}{rcl}
    \displaystyle{} R\I{wave}(\vartheta\I{LX}) &=& \displaystyle{} \dfrac{1}
      {P_{\mathrm{th}}(0)} \Int_{-\infty}^{\infty} \sigma(\vartheta\I{LX},t) \,
      J\I{X}(t) \differential t \\
    &=& \displaystyle{} \cos^2 \vartheta\I{LX} \, R_{\parallel / \mathrm{th}}(0)
      + \sin^2 \vartheta\I{LX} \, R_{\perp / \mathrm{th}}(0) \\
    &&\displaystyle{} + \sin \vartheta\I{LX} \, \cos \vartheta\I{LX} \,
      R_{\perp,\parallel / \mathrm{th}}(0)
  \end{array}
\end{equation}
is measured, varying~$\vartheta\I{LX}$.
In addition to Eq.~(\ref{eq:crossratios}), we define~$R_{\perp,\parallel
/ \mathrm{th}}(\tau) = \frac{P_{\perp,\parallel}(\tau)}{P_{\mathrm{th}}(\tau)}$ using
Eq.~(\ref{eq:crosscorr}) with~$\sigma_{\perp, \parallel}(t)$.

\section{Electronic structure model}
\label{sec:elstrmod}

\begin{figure}
  \begin{center}
    a)~~\includegraphics[clip,width=3cm]{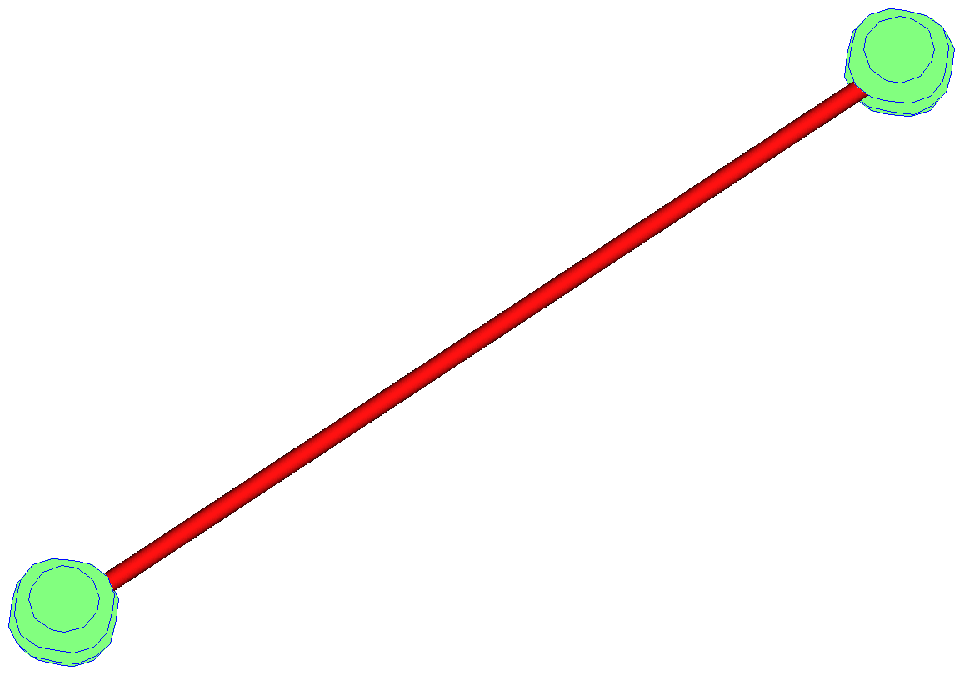}
    \qquad b)~\includegraphics[clip,width=3cm]{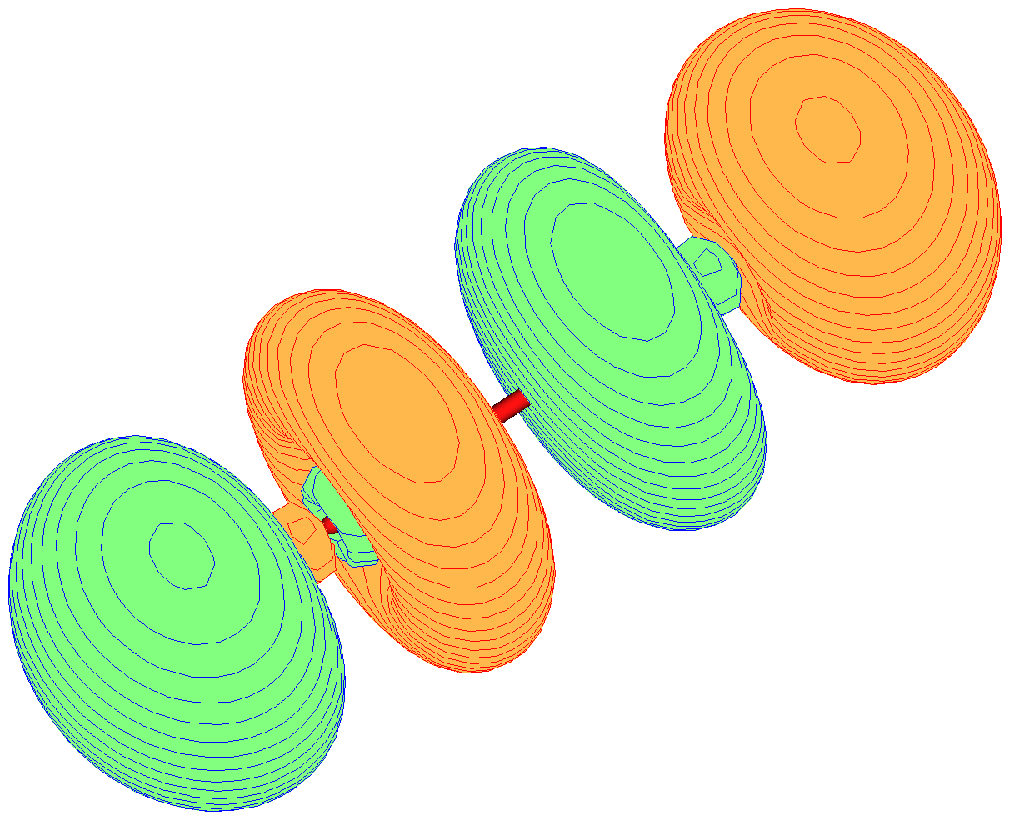}
    \caption{(Color online) The two orbitals included in the two-level model
             for the electronic structure of a bromine molecule from which
             the x-ray absorption cross section of the molecule is
             determined~\cite{footnote4}.\protect\nocite{Schaftenaar:MO-00}
             a)~Totally symmetric linear combination~$\sigma\I{g} \, 1s$
             ($\Sigma_g^+$~symmetry) of the two Br$\,1s$~orbitals;
             b)~the $\sigma\I{u} \, 4p$~antibonding molecular orbital of
             $\Sigma_u^+$~symmetry.}
    \label{fig:br2orbitals}
  \end{center}
\end{figure}

We assume a two-level model for the electronic structure
of a bromine molecule~(Br$_2$).
For simplicity, we treat the two states as Hartree-Fock
orbitals~\cite{Szabo:MQC-89};
yet our reasoning is general and not limited by this assumption.
The model is justified by the experimental $K$~edge absorption spectrum
of~Br$_2$ which is shown in Fig.~1 of Ref.~\onlinecite{Filipponi:AD-98};
it exhibits an isolated pre-edge resonance which is assigned to a transition
from the gerade linear combination of the Br$\,1s$~atomic orbitals, the
$\sigma\I{g} \, 1s$~molecular orbital, to the $\sigma\I{u} \, 4p$~antibonding molecular
orbital.
The two molecular orbitals are displayed in Fig.~\ref{fig:br2orbitals}
of this paper.

We need to find out what the dipole selection rules are for the
transition~$\sigma\I{g} \, 1s \to \sigma\I{u} \, 4p$ using
symmetry arguments~\cite{Bishop:GT-73,Atkins:MQM-04}.
Bromine forms linear molecules of~$D_{\infty h}$
symmetry~\cite{Atkins:MQM-04}.
The electronic transition is described by the matrix
element~$\vec d_{\sigma\I{g} \, 1s, \sigma\I{u} \, 4p} = \bra{\sigma\I{g} \, 1s}
\hat{\vec d} \ket{\sigma\I{u} \, 4p}$ in Eq.~(\ref{eq:diptransme}) formed with
the Cartesian dipole operator~$\hat{\vec d} = (a,b,c)\transpose$
in the molecule-fixed frame.
The vanishing integrals rule~\cite{Bishop:GT-73,Atkins:MQM-04} says that
the components of~$\vec d_{\sigma\I{g} \, 1s, \sigma\I{u} \, 4p}$ can only
be nonzero, if the direct product representation of the representations
to which~$\bra{\sigma\I{g} \, 1s}$, $\hat{\vec d}$, and
$\ket{\sigma\I{u} \, 4p}$ belong, contains the totally symmetric representation.
The question therefore is, for~$i \in \{a, b, c\}$, does
$\Gamma^{* \, \sigma\I{g} \, 1s} \otimes \Gamma^{i} \otimes
\Gamma^{\sigma\I{u} \, 4p}$ contain~$\Gamma^{\Sigma_g^+}$?

There are two linear combinations of the Br$\,1s$~orbitals.
These are of~$\Sigma_g^+$ and $\Sigma_u^+$~symmetry, respectively;
the $\sigma\I{u} \, 4p$~orbital has $\Sigma_u^+$~symmetry.
The Cartesian component~$c$ of the dipole operator is of $\Sigma_u^+$~symmetry
whereas $a$ and $b$ are of $\Pi_u$~symmetry.
With the help of the direct product table of~$D_{\infty
h}$~\cite{Atkins:MQM-04}, we find that only the direct product of the combination of the
Br$\,1s$~orbitals with~$\Sigma_g^+$ symmetry with the $c$~component of~$\hat{\vec d}$
and $\sigma\I{u} \, 4p$ leads to a non-vanishing transition matrix element.
This linear combination is thus identified with the $\sigma\I{g} \, 1s$~molecular
orbital in the two-level model.
Only the $c$~component~$d_c$ of~$\vec d_{\sigma\I{g} \, 1s, \sigma\I{u} \, 4p}$
is nonvanishing.
This implies that, according to the scalar product in Eq.~(\ref{eq:H_X_matel}),
x~rays are only absorbed with an x-ray polarization vector
component along the internuclear axis;
there is no absorption for perpendicular polarizations.
In computations, we focus on ratios such as Eq.~(\ref{eq:crossratios}).
Hence, our results are independent of the value of~$d_c$.
Since the molecular orbitals are glued to the rigid nuclear framework,
they rotate with the molecule.
Hence, the photoabsorption is sensitive to the spatial alignment
of the molecules.

\newcommand{\crulefill}{\hrulefill}

\begin{table*}
  \centering
  \begin{ruledtabular}
    \begin{tabular}{rlcccccc}
                &     & \multicolumn{2}{c}{\crulefill CCS\crulefill}
                      & \multicolumn{2}{c}{\crulefill CC2\crulefill}
                      & \multicolumn{2}{c}{\crulefill CCSD\crulefill} \\
    \hfill  Basis set\hfill\hfill & $\omega\I{L}$ [$\Hartree$] & $\bar\alpha$ [a.u.]
      & $\Delta \alpha$ [a.u.] & $\bar\alpha$ [a.u.] & $\Delta \alpha$ [a.u.]
      & $\bar\alpha$ [a.u.] & $\Delta \alpha$ [a.u.] \\
      \hline
          cc-pVDZ & 0     & 30.87 & 37.76 & 28.03 & 33.91 & 27.51 & 32.40 \\
                  & 0.057 & 31.28 & 38.82 & 28.43 & 34.86 & 27.89 & 33.31 \\[3pt]
      aug-cc-pVDZ & 0     & 45.00 & 30.99 & 43.03 & 27.85 & 42.31 & 26.47 \\
                  & 0.057 & 45.68 & 31.89 & 43.79 & 28.60 & 43.04 & 27.19 \\[3pt]
      aug-cc-pVTZ & 0     & 47.31 & 30.66 & 46.03 & 27.57 & 44.96 & 26.41 \\
                  & 0.057 & 48.06 & 31.52 & 46.87 & 28.30 & 45.76 & 27.11 \\[3pt]
      aug-cc-pVQZ & 0     & 47.50 & 30.72 & 46.13 & 27.74 & 44.98 & 26.73 \\
                  & 0.057 & 48.26 & 31.58 & 46.99 & 28.48 & 45.79 & 27.45
    \end{tabular}
  \end{ruledtabular}
  \caption{Average dynamical polarizabilities~$\bar\alpha(\omega\I{L})$ and
           dynamical polarizability anisotropies~$\Delta \alpha(\omega\I{L})$
           of a bromine molecule in the light of a laser with photon
           energy~$\omega\I{L}$.
           The polarizabilities are determined with coupled-cluster linear
           response methods for several basis sets.
           Theoretical SDQ-MP4 values, $\bar\alpha\I{ref}(0)
           = 44.79 \U{a.u.}$ and $\Delta \alpha\I{ref}(0) = 25.68 \U{a.u.}$,
           were taken from Table~3 in Ref.~\onlinecite{Archibong:SP-93}.}
  \label{tab:polarizability}
\end{table*}

For the two-level model, one can further simplify the dependence on the
angle between laser and x-ray polarizations~$\vartheta\I{LX}$ of the
cross section~(\ref{eq:angxsect}).
Due to the fact that the dipole vector~$\vec d_{\sigma\I{g} \, 1s, \sigma\I{u}
\, 4p}$ has only a nonvanishing $c$~component (or $m=0$~component)~$d_c$, the
cross term in Eq.~(\ref{eq:angxsectdep}) becomes
\begin{equation}
  \label{eq:angxsectcross}
  \begin{array}{cl}
    &|d_c|^2 \, \sin \vartheta\I{LX} \, \cos \vartheta\I{LX} \,
      [   s_{c, J'KM, J''K''M''}^{\>\prime, \perp} \, s_{c, J''K''M'',
      JKM}^{\>\prime, \parallel} \\
    &\displaystyle{} +  s_{c, J'KM, J''K''M''}^{\>\prime, \parallel} \,
      s_{c, J''K''M'', JKM}^{\>\prime, \perp} ] \; ,
  \end{array}
\end{equation}
letting~$s_{c, J'KM, J''K''M''}^{\>\prime, \perp} \equiv
(\vec s_{J'KM, J''K''M''}^{\>\prime, \perp})_0$
and $s_{c, J'KM, J''K''M''}^{\>\prime, \parallel} \equiv
(\vec s_{J'KM, J''K''M''}^{\>\prime, \parallel})_0$.
Recalling the rule~(\ref{eq:3D-rule}) to evaluate the matrix
elements in Eq.~(\ref{eq:angxsectcross}), we obtain conditions that need
to be fulfilled such that the Clebsch-Gordan coefficients do not vanish.
As we focus on the $m=0$~component, we obtain from Eq.~(\ref{eq:3D-rule})
the equality~$K = K''$.
Further, for~$s_{c, J'KM, J''K''M''}^{\>\prime, \perp}$ to be nonzero,
$M = M'' \pm 1$ must hold whereas for~$s_{c, J'KM, J''K''M''}^{\>\prime,
\parallel} \neq 0$ the assertion~$M = M''$ is necessary.
Both conditions cannot be fulfilled simultaneously such that the cross term
in Eq.~(\ref{eq:angxsectcross}) and, therefore, $\sigma^{\perp,
\parallel}(t)$ in Eq.~(\ref{eq:angxsect}) vanishes.
Our simplifications carry over to the signal of a wave plate
scan;
in Eq.~(\ref{eq:waveplatefull}), $R_{\perp,\parallel / \mathrm{th}}(0) = 0$ and thus
\begin{equation}
  \label{eq:waveplate}
  R\I{wave, model}(\vartheta\I{LX}) = \cos^2 \vartheta\I{LX} \, R_{\parallel /
  \mathrm{th}}(0) + \sin^2 \vartheta\I{LX} \, R_{\perp / \mathrm{th}}(0) \; ,
\end{equation}
with Eq.~(\ref{eq:crossratios}).

A two-level electronic structure model, similar to the one used in this section,
is suitable when the x-ray energy lies on an isolated resonance in the
absorption spectrum of a molecule.
Neglecting the coupling between nuclear and electronic motion
as well as vibrations due to electronic excitation~\cite{Thomas:ANL-02}, as
we have done in this paper, the two-level model can easily be generalized to a full
\emph{ab initio} description because, in this case, the electronic processes
are completely decoupled from the nuclear motion.
The electronic transitions matrix elements are determined in
the fixed-nuclei approximation and inserted into Eq.~(\ref{eq:xsect}).

\section{Computational details}
\label{sec:compdet}

Based on the theory of Secs.~\ref{sec:theory} and \ref{sec:elstrmod},
we have developed the program \textsc{alignmol} of the \textsc{fella}
package~\cite{fella:pgm-V1.2.0}.
The program handles symmetric-top molecules where the special case of
a linear molecule is treated by letting~$K = 0$ in all equations of this paper
and eliminating sums over~$K$.
A large number of parameters are needed to specify the molecular properties,
the numerical accuracy, and the laser- and x-ray pulses which we discuss
in the following paragraphs.

The Br$_2$ molecule is a linear rotor with an internuclear distance
of~$R = 2.289 \angstrom$~\cite{Filipponi:AD-98}.
The rotational constants in Eq.~(\ref{eq:syrotev}) are given by~$A=0$
and $B = \frac{1}{2 \mu R^2}$, where $\mu$~is the reduced
mass~\cite{Kroto:MR-75}.
There are two stable, naturally occurring isotopes of bromine: $^{79}$Br and
$^{81}$Br~\cite{Mills:QU-88}.
The first one, $^{79}$Br, has a mass of $78.9183361 \U{u}$ and a natural
abundance of~$p_{\>^{79}\mathrm{Br}} = 50.69\%$ whereas the second one,
$^{81}$Br, has a mass of~$80.916289 \U{u}$ and a natural abundance
of~$p_{\>^{81}\mathrm{Br}} = 49.31\%$~\cite{Mills:QU-88,Rosman:IC-98}.
This leads to three possible combinations of isotopes in~Br$_2$ and thus to
three different rotational constants:
$B = 0.081537 \invcm$~for $^{79}$Br$_2$, $B = 0.079524 \invcm$
for~$^{81}$Br$_2$, and $B = 0.080530 \invcm$~for $^{79}$Br--$^{81}$Br.

Both bromine isotopes have a nuclear spin quantum number
of~$I = \frac{3}{2}$~\cite{Mills:QU-88}.
The statistical weights in Eq.~(\ref{eq:thdenseig}) for an isotope-pure
diatomic molecule with nuclear spin~$I$, are given
by~\cite{Townes:MS-55,Kroto:MR-75}
\begin{equation}
  g_I(J,K) = \cases{ I + 1 & ; $J$~even \cr
                     I     & ; $J$~odd } \; .
\end{equation}
For $^{79}$Br--$^{81}$Br, the nuclei are distinguishable and the statistical
weight is~$g_I(J,K) \equiv 1$.
As the equations in this paper are mostly nonlinear in the rotational
constants and the statistical weights, it is required that for each of the
three combinations of bromine isotopes, a separate computation of the
desired quantities is carried out.
The results are then averaged assigning the
weights~$p_{\>^{79}\mathrm{Br}}^2$ to results for~$^{79}$Br$_2$,
$p_{\>^{81}\mathrm{Br}}^2$ to~$^{81}$Br$_2$, and $2 \, p_{\>^{79}\mathrm{Br}}
p_{\>^{81}\mathrm{Br}}$ to~$^{79}$Br--$^{81}$Br.

We assume a rotational temperature of~Br$_2$ of~$T = 30 \U{K}$;
the initial rotational density matrix~(\ref{eq:thdenseig}) needs
to be represented in terms of a finite number of symmetric-rotor
functions~(\ref{eq:syrofunc}) [for $K = 0$ because Br$_2$ is linear].
We chose angular momenta up to~$J\I{max} = 70$.
The value for~$J\I{max}$ is determined to be sufficiently large such that the
contribution with~$J = J\I{max}$ to the partition function~(\ref{eq:partfunc})
is~$< 10^{-6}$.
For each~$J = 0, \ldots, J\I{max}$ all symmetric-rotor
functions~(\ref{eq:syrofunc}) with~$M = -J, \ldots, J$ are employed
to represent the density operator;
sums over~$K$ are omitted.
The total energy of the Gaussian laser pulse~(\ref{eq:GaussEnv})
is~$E\I{pulse} = 2 \U{mJ}$;
it lasts (FWHM)~$\tau\I{L} = 100 \U{ps}$.
The energy and duration of the pulse imply a peak intensity
of~$\approx 10^{12} \U{\frac{W}{cm^2}}$ on the axis of a laser beam with a
Gaussian radial profile perpendicular to the beam axis of a FWHM width
of~$\varrho\I{L} = 40 \U{\mu m}$.
We do not consider a radial intensity variation but take the value on
axis.

As the molecule is in the field of the aligning laser, we need to
determine the dynamical polarizability tensor in the molecule-fixed
reference frame~$\mat \alpha^{\mathrm M}(\omega\I{L})$~\cite{Bishop:MV-90,%
Christiansen:ID-98}.
This means specifically the average dynamical
polarizability~$\bar\alpha(\omega\I{L})$ [Eq.~(\ref{eq:polave})] and the
dynamical polarizability anisotropy~$\Delta \alpha(\omega\I{L})$
[Eq.~(\ref{eq:polani})] which both enter Eq.~(\ref{eq:lasermat}).
The quantities can be computed in fixed-nuclei approximation utilizing the
\emph{ab initio} quantum chemistry program package
\textsc{dalton}~\cite{dalton:pgm-05}.
We employ molecular linear-response theory in conjunction with
wave function models of increasing sophistication:
coupled-cluster singles~(CCS), second-order approximate coupled-cluster
singles and doubles~(CC2), and coupled-cluster singles and
doubles~(CCSD)~\cite{Christiansen:ID-98}.
The cc-pVDZ [correlation consistent polarized valence double zeta] and
aug-cc-pV$X$Z [augmented cc-pV$X$Z] (for $X={}$D, T, Q [D:~double,
T:~triple, Q:~quadruple]) basis
sets~\cite{Wilson:GBS-99,basislib:02-02-06} are used to describe the
bromine atoms.
We examine the static case, $\omega\I{L} = 0 \Hartree$, and the dynamic case for the
photon energy of an $800 \U{nm}$ Ti:Sapphire laser, $\omega\I{L} = 0.057 \Hartree$.

Our average polarizabilities and the polarizability anisotropies
are listed in Table~\ref{tab:polarizability}.
A moderately rapid convergence with respect to the basis set quality is
observed for the individual methods;
the difference is much smaller between corresponding results
for~aug-cc-pVTZ and aug-cc-pVQZ than between the results for~$\omega\I{L}
= 0 \Hartree$ and $\omega\I{L} = 0.057 \Hartree$,
\ie, the basis set limit is nearly reached.
We conclude that the remaining deviations between methods for a
specific~$\omega\I{L}$ and the aug-cc-pVQZ~basis reflect the differences
between the coupled-cluster methods.
Comparing our best results for~$\omega\I{L} = 0 \Hartree$, CCSD and aug-cc-pVQZ,
with theoretical SDQ-MP4 values from Table~3 in
Ref.~\onlinecite{Archibong:SP-93}, we note a satisfactory agreement.
In our computations, we use the value of~$\Delta \alpha(0.057 \Hartree)
= 27.45 \U{a.u.}$ for the polarizability anisotropy.

\begin{figure}
  \begin{center}
    \includegraphics[clip,width=\hsize]{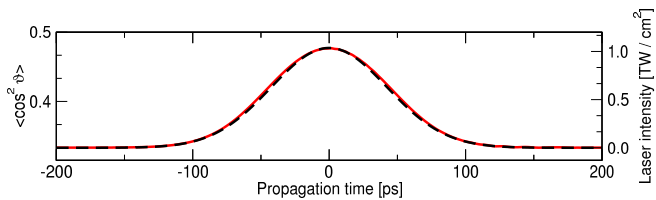}
    \caption{(Color online) The time evolution of~$\expectval{\cos^2 \vartheta}(t)$
             [Eq.~(\ref{eq:cos2theta})] of bromine molecules is given by the solid
             (red) line as a function of the laser pulse which is depicted by a
             dashed (black) line.}
    \label{fig:cosine}
  \end{center}
\end{figure}

The coefficients~(\ref{eq:wfcoeff}) of the rotational density
matrix in Eq.~(\ref{eq:densnuclprop}) are represented here by~$N\I{eq}
= J\I{max}$ values for~$J'$ which determines the number of equations of
motion~(\ref{eq:coeffEOM}) for each initial symmetric-rotor state~$\ket{JKM}$.
We use the Runge-Kutta algorithm of fourth order~\cite{Golub:SC-92}
to compute the time evolution of the density matrix via
Eq.~(\ref{eq:coeffEOM}).
The propagation takes place over the time interval~$[-400 \U{ps}, 400 \U{ps}]$
which is subdivided into~$N\I{time} = 6002$ time steps.

A Gaussian shape of the x-ray pulses similar to the laser
pulse~(\ref{eq:GaussEnv}) is assumed.
As the interaction with the x~rays is treated in terms of a one-photon
process, the specification of a value for the peak intensity becomes
unnecessary;
the relevant prefactor cancels once ratios are formed as
in Eqs.~(\ref{eq:crossratios}) and (\ref{eq:waveplate}).
Only, the FWHM duration of the x-ray pulse needs to be given.
We choose~$\tau\I{X} = 120 \U{ps}$, a typical value for a
third-generation synchrotron radiation source~\cite{Thompson:XR-01}.
The cross correlation of the cross section and the x-ray
flux~(\ref{eq:crosscorr}) is computed via numerical
integration using the trapezoid rule~\cite{Golub:SC-92}.
To this end, we assume the laser pulse to be centered at zero,
Eq.~(\ref{eq:GaussEnv}), and use an equidistant grid of~$N\I{X} = 1003$
x-ray pulses, \ie, 1003~different time delays~$\tau$, over the whole
propagation interval~$[-400 \U{ps}, 400 \U{ps}]$.

\section{Results and discussion}
\label{sec:results}

Here, we exemplify the theory of the previous sections for bromine
molecules.
We picked laser and x-ray pulse parameters in the previous
Sec.~\ref{sec:compdet};
the pulse durations were chosen such that they are comparable;
the resulting molecular dynamics is shown in Fig.~\ref{fig:cosine}.
We give the time evolution of~$\expectval{\cos^2 \vartheta}(t)$
for the rotational wave packet [Eq.~(\ref{eq:cos2theta})].
From the figure it becomes clear that we are in the regime
of adiabatic alignment because the temporal evolution
of~$\expectval{\cos^2 \vartheta}(t)$ and of the laser pulse are practically
the same~\cite{Stapelfeldt:AM-03,Seideman:NA-05}.
Even the shape of the molecular response is very similar to the one
of the laser pulse.
Yet the response of the molecule is not totally Gaussian as
indicated by the slight deviation between the two curves in
Fig.~\ref{fig:cosine}.

\begin{figure}
  \begin{center}
    \includegraphics[clip,width=\hsize]{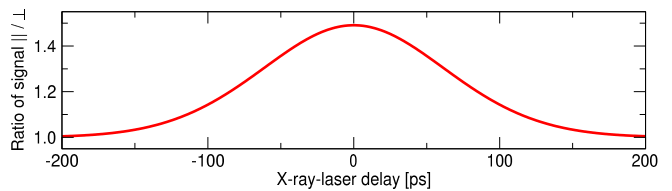}
    \caption{(Color online) Ratio of cross correlations~$R\I{\parallel /
             \perp}(\tau)$ [Eq.~(\ref{eq:crossparperp})]
             for bromine molecules as a function of the time delay
             between laser and x-ray pulses.}
    \label{fig:crosscorr}
  \end{center}
\end{figure}

The ratio of cross correlations~$R\I{\parallel / \perp}(\tau)$
[Eq.~(\ref{eq:crossparperp})] between cross section and x-ray flux
is displayed in Fig.~\ref{fig:crosscorr}.

\begin{figure}
  \begin{center}
    \includegraphics[clip,width=\hsize]{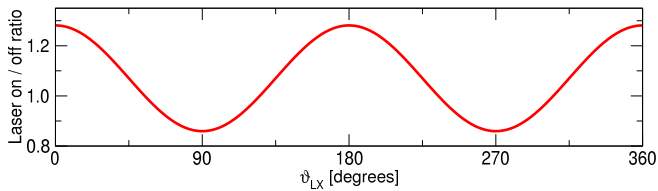}
    \caption{(Color online) Dependence of the ratio between the x-ray
             absorption signals of laser-aligned bromine molecules versus a
             thermal ensemble on the angle~$\vartheta_{\mathrm LX}$ between
             laser and x-ray polarization vectors.}
    \label{fig:waveplate}
  \end{center}
\end{figure}

The ratio of total x-ray absorption between laser on and laser off is given
by Eq.~(\ref{eq:waveplate}).
It is plotted in Fig.~\ref{fig:waveplate} in relation to the angle between
laser and x-ray polarization vectors~$\vartheta\I{LX}$.
Without laser, there is only a thermal ensemble and the angular dependence
vanishes.

\section{Conclusion}
\label{sec:conclusion}

We devised a theory for the x-ray absorption of symmetric-top molecules
subject to an intense nonresonant laser.
The rigid-rotor approximation was assumed and a density-matrix
formalism was used to describe a thermal ensemble of rotational
states evolving in time.
The x-ray absorption is treated in terms of a one-photon process.
The very different time scales of molecular rotation and electronic
relaxation (core-hole lifetime) facilitated a separation of the
two problems, \ie, the molecular rotation was treated separately
from the x-ray absorption.
The electronic structure of bromine molecules was described in terms of
a simple parameter-free two-level model for the $\sigma\I{g} \, 1s \to
\sigma\I{u} \, 4p$~pre-edge resonance.
The symmetry of the resonance caused a linear dichroism---light
was only absorbed for a linear polarization component along the
internuclear axis of~Br$_2$---which lead to a modulation of the
photoabsorption with respect to the alignment of the molecule.

Our theory was realized in terms of the \textsc{alignmol} program which is
part of the \textsc{fella} package~\cite{fella:pgm-V1.2.0}.
The dynamic dipole polarizability for the laser frequency
was determined for~Br$_2$ using three \emph{ab initio}
coupled-cluster methods of increasing precision: CCS, CC2, and CCSD;
it was found not to deviate significantly from the static polarizability.
We investigated the time evolution of the alignment of Br$_2$~molecules
in the adiabatic regime.
The x-ray absorption is studied in terms of the experimentally
accessible ratio of cross correlations between cross section and x-ray pulse
for several combinations of laser on and off and parallel and perpendicular
linear polarization vectors of the laser and the x~rays.
A continuous variation of the angle between the polarization
vectors was also investigated.

This article forms the theoretical basis for future studies.
In forthcoming papers, we will apply our theory
to the symmetric-top molecule
bromotrifluoromethane~(\CFtBr{})~\cite{Santra:SF-07,Peterson:XR-up}.
There, we will compare our results with experimental data.
Further perspectives include obtaining absolute cross sections from
our theory.
For this purpose one needs a quantitative
description of the electronic structure of the molecule, \ie,
one requires transition dipole moments instead of our two-level
electronic structure model.
The moments can be obtained from existing \emph{ab initio} program packages.
Our work provides an alternative way to the established but
intricate~\cite{Pavicic:DM-07} Coulomb-explosion
technique~\cite{Stapelfeldt:AM-03} to study laser-controlled molecular
rotation in the gas phase which additionally opens up the possibility to
study the laser-manipulated rotation of an ensemble of molecules in
solution~\cite{Ramakrishna:IL-05} where the
Coulomb-explosion technique is bound to fail.
Moreover, with our theory, one can interpret experimental data;
the symmetry of molecular resonances in the gas phase enters our formulas
in terms of selection rules for dipole matrix elements.
In conjunction with experiments, our approach offers a novel route to
determine the symmetry of resonances.
It thus complements conventional methods using clamped molecules on
surfaces~\cite{Stohr:NE-96} and angle-resolved photoion yield
spectroscopy~\cite{Adachi:SR-05}.

\begin{acknowledgments}
We would like to thank Linda Young and Phay Ho for fruitful discussions and
a critical reading of the manuscript.
C.B.'s and R.S.'s work was supported by the Office of Basic Energy Sciences,
Office of Science, U.S.~Department of Energy, under Contract
No.~DE-AC02-06CH11357.
C.B.'s research was partly funded by a Feodor Lynen Research Fellowship from
the Alexander von Humboldt Foundation.
\end{acknowledgments}


\begin{thebibliography}{54}
\expandafter\ifx\csname natexlab\endcsname\relax\def\natexlab#1{#1}\fi
\expandafter\ifx\csname bibnamefont\endcsname\relax
  \def\bibnamefont#1{#1}\fi
\expandafter\ifx\csname bibfnamefont\endcsname\relax
  \def\bibfnamefont#1{#1}\fi
\expandafter\ifx\csname citenamefont\endcsname\relax
  \def\citenamefont#1{#1}\fi
\expandafter\ifx\csname url\endcsname\relax
  \def\url#1{\texttt{#1}}\fi
\expandafter\ifx\csname urlprefix\endcsname\relax\def\urlprefix{URL }\fi
\providecommand{\bibinfo}[2]{#2}
\providecommand{\eprint}[2][]{\url{#2}}

\bibitem[{\citenamefont{Santra et~al.}(2006)\citenamefont{Santra, Dunford, and
  Young}}]{Santra:SO-06}
\bibinfo{author}{\bibfnamefont{R.}~\bibnamefont{Santra}},
  \bibinfo{author}{\bibfnamefont{R.~W.} \bibnamefont{Dunford}},
  \bibnamefont{and} \bibinfo{author}{\bibfnamefont{L.}~\bibnamefont{Young}},
  \bibinfo{journal}{Phys. Rev. A} \textbf{\bibinfo{volume}{74}},
  \bibinfo{pages}{043403} (\bibinfo{year}{2006}).

\bibitem[{\citenamefont{Loh et~al.}(2007)\citenamefont{Loh, Khalil, Correa,
  Santra, Buth, and Leone}}]{Loh:QS-07}
\bibinfo{author}{\bibfnamefont{Z.-H.} \bibnamefont{Loh}},
  \bibinfo{author}{\bibfnamefont{M.}~\bibnamefont{Khalil}},
  \bibinfo{author}{\bibfnamefont{R.~E.} \bibnamefont{Correa}},
  \bibinfo{author}{\bibfnamefont{R.}~\bibnamefont{Santra}},
  \bibinfo{author}{\bibfnamefont{C.}~\bibnamefont{Buth}}, \bibnamefont{and}
  \bibinfo{author}{\bibfnamefont{S.~R.} \bibnamefont{Leone}},
  \bibinfo{journal}{Phys. Rev. Lett.} \textbf{\bibinfo{volume}{98}},
  \bibinfo{pages}{143601} (\bibinfo{year}{2007}),
  \bibinfo{note}{\href{http://arxiv.org/abs/physics/0703149}
  {arXiv:physics/0703149}}.

\bibitem[{\citenamefont{Santra et~al.}(2007)\citenamefont{Santra, Buth,
  Peterson, Dunford, Kanter, Kr\"assig, Southworth, and Young}}]{Santra:SF-07}
\bibinfo{author}{\bibfnamefont{R.}~\bibnamefont{Santra}},
  \bibinfo{author}{\bibfnamefont{C.}~\bibnamefont{Buth}},
  \bibinfo{author}{\bibfnamefont{E.~R.} \bibnamefont{Peterson}},
  \bibinfo{author}{\bibfnamefont{R.~W.} \bibnamefont{Dunford}},
  \bibinfo{author}{\bibfnamefont{E.~P.} \bibnamefont{Kanter}},
  \bibinfo{author}{\bibfnamefont{B.}~\bibnamefont{Kr\"assig}},
  \bibinfo{author}{\bibfnamefont{S.~H.} \bibnamefont{Southworth}},
  \bibnamefont{and} \bibinfo{author}{\bibfnamefont{L.}~\bibnamefont{Young}},
  \bibinfo{journal}{J. Phys.: Conf. Ser.} \textbf{\bibinfo{volume}{88}},
  \bibinfo{pages}{012052} (\bibinfo{year}{2007}),
  \bibinfo{note}{\href{http://arxiv.org/abs/0712.2556} {arXiv:0712.2556}}.

\bibitem[{\citenamefont{Buth and Santra}(2007{\natexlab{a}})}]{Buth:TX-07}
\bibinfo{author}{\bibfnamefont{C.}~\bibnamefont{Buth}} \bibnamefont{and}
  \bibinfo{author}{\bibfnamefont{R.}~\bibnamefont{Santra}},
  \bibinfo{journal}{Phys. Rev. A} \textbf{\bibinfo{volume}{75}},
  \bibinfo{pages}{033412} (\bibinfo{year}{2007}{\natexlab{a}}),
  \bibinfo{note}{\href{http://arxiv.org/abs/physics/0611122}
  {arXiv:physics/0611122}}.

\bibitem[{\citenamefont{Buth et~al.}(2007)\citenamefont{Buth, Santra, and
  Young}}]{Buth:ET-07}
\bibinfo{author}{\bibfnamefont{C.}~\bibnamefont{Buth}},
  \bibinfo{author}{\bibfnamefont{R.}~\bibnamefont{Santra}}, \bibnamefont{and}
  \bibinfo{author}{\bibfnamefont{L.}~\bibnamefont{Young}},
  \bibinfo{journal}{Phys. Rev. Lett.} \textbf{\bibinfo{volume}{98}},
  \bibinfo{pages}{253001} (\bibinfo{year}{2007}),
  \bibinfo{note}{\href{http://arxiv.org/abs/0705.3615} {arXiv:0705.3615}}.

\bibitem[{\citenamefont{Normand et~al.}(1992)\citenamefont{Normand, Lompre, and
  Cornaggia}}]{Normand:LI-92}
\bibinfo{author}{\bibfnamefont{D.}~\bibnamefont{Normand}},
  \bibinfo{author}{\bibfnamefont{L.~A.} \bibnamefont{Lompre}},
  \bibnamefont{and}
  \bibinfo{author}{\bibfnamefont{C.}~\bibnamefont{Cornaggia}},
  \bibinfo{journal}{J. Phys. B} \textbf{\bibinfo{volume}{25}},
  \bibinfo{pages}{L497} (\bibinfo{year}{1992}).

\bibitem[{\citenamefont{Friedrich and Herschbach}(1995)}]{Friedrich:AT-95}
\bibinfo{author}{\bibfnamefont{B.}~\bibnamefont{Friedrich}} \bibnamefont{and}
  \bibinfo{author}{\bibfnamefont{D.}~\bibnamefont{Herschbach}},
  \bibinfo{journal}{Phys. Rev. Lett.} \textbf{\bibinfo{volume}{74}},
  \bibinfo{pages}{4623} (\bibinfo{year}{1995}).

\bibitem[{\citenamefont{Stapelfeldt and Seideman}(2003)}]{Stapelfeldt:AM-03}
\bibinfo{author}{\bibfnamefont{H.}~\bibnamefont{Stapelfeldt}} \bibnamefont{and}
  \bibinfo{author}{\bibfnamefont{T.}~\bibnamefont{Seideman}},
  \bibinfo{journal}{Rev. Mod. Phys.} \textbf{\bibinfo{volume}{75}},
  \bibinfo{pages}{543} (\bibinfo{year}{2003}).

\bibitem[{\citenamefont{Seideman and Hamilton}(2005)}]{Seideman:NA-05}
\bibinfo{author}{\bibfnamefont{T.}~\bibnamefont{Seideman}} \bibnamefont{and}
  \bibinfo{author}{\bibfnamefont{E.}~\bibnamefont{Hamilton}}, in
  \emph{\bibinfo{booktitle}{Advances in atomic, molecular, and optical
  physics}}, edited by \bibinfo{editor}{\bibfnamefont{P.~R.}
  \bibnamefont{Berman}} \bibnamefont{and} \bibinfo{editor}{\bibfnamefont{C.~C.}
  \bibnamefont{Lin}} (\bibinfo{publisher}{Elsevier},
  \bibinfo{address}{Amsterdam}, \bibinfo{year}{2005}),
  vol.~\bibinfo{volume}{52}, pp. \bibinfo{pages}{289--329}.

\bibitem[{\citenamefont{Ortigoso et~al.}(1999)\citenamefont{Ortigoso,
  Rodr\'iguez, Gupta, and Friedrich}}]{Ortigoso:TE-99}
\bibinfo{author}{\bibfnamefont{J.}~\bibnamefont{Ortigoso}},
  \bibinfo{author}{\bibfnamefont{M.}~\bibnamefont{Rodr\'iguez}},
  \bibinfo{author}{\bibfnamefont{M.}~\bibnamefont{Gupta}}, \bibnamefont{and}
  \bibinfo{author}{\bibfnamefont{B.}~\bibnamefont{Friedrich}},
  \bibinfo{journal}{J. Chem. Phys.} \textbf{\bibinfo{volume}{110}},
  \bibinfo{pages}{3870} (\bibinfo{year}{1999}).

\bibitem[{\citenamefont{Hamilton et~al.}(2005)\citenamefont{Hamilton, Seideman,
  Ejdrup, Poulsen, Bisgaard, Viftrup, and Stapelfeldt}}]{Hamilton:AS-05}
\bibinfo{author}{\bibfnamefont{E.}~\bibnamefont{Hamilton}},
  \bibinfo{author}{\bibfnamefont{T.}~\bibnamefont{Seideman}},
  \bibinfo{author}{\bibfnamefont{T.}~\bibnamefont{Ejdrup}},
  \bibinfo{author}{\bibfnamefont{M.~D.} \bibnamefont{Poulsen}},
  \bibinfo{author}{\bibfnamefont{C.~Z.} \bibnamefont{Bisgaard}},
  \bibinfo{author}{\bibfnamefont{S.~S.} \bibnamefont{Viftrup}},
  \bibnamefont{and}
  \bibinfo{author}{\bibfnamefont{H.}~\bibnamefont{Stapelfeldt}},
  \bibinfo{journal}{Phys. Rev. A} \textbf{\bibinfo{volume}{72}},
  \bibinfo{pages}{043402} (\bibinfo{year}{2005}).

\bibitem[{\citenamefont{Thompson et~al.}(2001)\citenamefont{Thompson, Attwood,
  Gullikson, Howells, Kortright, Robinson, Underwood, Kim, Kirz, Lindau
  et~al.}}]{Thompson:XR-01}
\bibinfo{author}{\bibfnamefont{A.~C.} \bibnamefont{Thompson}},
  \bibinfo{author}{\bibfnamefont{D.~T.} \bibnamefont{Attwood}},
  \bibinfo{author}{\bibfnamefont{E.~M.} \bibnamefont{Gullikson}},
  \bibinfo{author}{\bibfnamefont{M.~R.} \bibnamefont{Howells}},
  \bibinfo{author}{\bibfnamefont{J.~B.} \bibnamefont{Kortright}},
  \bibinfo{author}{\bibfnamefont{A.~L.} \bibnamefont{Robinson}},
  \bibinfo{author}{\bibfnamefont{J.~H.} \bibnamefont{Underwood}},
  \bibinfo{author}{\bibfnamefont{K.-J.} \bibnamefont{Kim}},
  \bibinfo{author}{\bibfnamefont{J.}~\bibnamefont{Kirz}},
  \bibinfo{author}{\bibfnamefont{I.}~\bibnamefont{Lindau}},
  \bibnamefont{et~al.}, \emph{\bibinfo{title}{X-ray data booklet}}
  (\bibinfo{publisher}{Lawrence Berkeley National Laboratory},
  \bibinfo{address}{Berkeley}, \bibinfo{year}{2001}), \bibinfo{edition}{2nd}
  ed.

\bibitem[{\citenamefont{Rehr and Albers}(2000)}]{Rehr:TA-00}
\bibinfo{author}{\bibfnamefont{J.~J.} \bibnamefont{Rehr}} \bibnamefont{and}
  \bibinfo{author}{\bibfnamefont{R.~C.} \bibnamefont{Albers}},
  \bibinfo{journal}{Rev. Mod. Phys.} \textbf{\bibinfo{volume}{72}},
  \bibinfo{pages}{621} (\bibinfo{year}{2000}).

\bibitem[{\citenamefont{Als-Nielsen and McMorrow}(2001)}]{Als-Nielsen:EM-01}
\bibinfo{author}{\bibfnamefont{J.}~\bibnamefont{Als-Nielsen}} \bibnamefont{and}
  \bibinfo{author}{\bibfnamefont{D.}~\bibnamefont{McMorrow}},
  \emph{\bibinfo{title}{Elements of modern x-ray physics}}
  (\bibinfo{publisher}{John Wiley~\& Sons}, \bibinfo{address}{New York},
  \bibinfo{year}{2001}), ISBN \bibinfo{isbn}{0-471-49858-0}.

\bibitem[{\citenamefont{Haack et~al.}(2000)\citenamefont{Haack, Ceballos,
  Wende, Baberschke, Arvanitis, Ankudinov, and Rehr}}]{Haack:SR-00}
\bibinfo{author}{\bibfnamefont{N.}~\bibnamefont{Haack}},
  \bibinfo{author}{\bibfnamefont{G.}~\bibnamefont{Ceballos}},
  \bibinfo{author}{\bibfnamefont{H.}~\bibnamefont{Wende}},
  \bibinfo{author}{\bibfnamefont{K.}~\bibnamefont{Baberschke}},
  \bibinfo{author}{\bibfnamefont{D.}~\bibnamefont{Arvanitis}},
  \bibinfo{author}{\bibfnamefont{A.~L.} \bibnamefont{Ankudinov}},
  \bibnamefont{and} \bibinfo{author}{\bibfnamefont{J.~J.} \bibnamefont{Rehr}},
  \bibinfo{journal}{Phys. Rev. Lett.} \textbf{\bibinfo{volume}{84}},
  \bibinfo{pages}{614} (\bibinfo{year}{2000}).

\bibitem[{\citenamefont{Pavi\v{c}i\'c et~al.}(2007)\citenamefont{Pavi\v{c}i\'c,
  Lee, Rayner, Corkum, and Villeneuve}}]{Pavicic:DM-07}
\bibinfo{author}{\bibfnamefont{D.}~\bibnamefont{Pavi\v{c}i\'c}},
  \bibinfo{author}{\bibfnamefont{K.~F.} \bibnamefont{Lee}},
  \bibinfo{author}{\bibfnamefont{D.~M.} \bibnamefont{Rayner}},
  \bibinfo{author}{\bibfnamefont{P.~B.} \bibnamefont{Corkum}},
  \bibnamefont{and} \bibinfo{author}{\bibfnamefont{D.~M.}
  \bibnamefont{Villeneuve}}, \bibinfo{journal}{Phys. Rev. Lett.}
  \textbf{\bibinfo{volume}{98}}, \bibinfo{pages}{243001}
  (\bibinfo{year}{2007}).

\bibitem[{\citenamefont{Campbell and Papp}(2001)}]{Campbell:WA-01}
\bibinfo{author}{\bibfnamefont{J.~L.} \bibnamefont{Campbell}} \bibnamefont{and}
  \bibinfo{author}{\bibfnamefont{T.}~\bibnamefont{Papp}}, \bibinfo{journal}{At.
  Data Nucl. Data Tables} \textbf{\bibinfo{volume}{77}}, \bibinfo{pages}{1}
  (\bibinfo{year}{2001}).

\bibitem[{\citenamefont{Peterson et~al.}(2008)\citenamefont{Peterson, Buth,
  Arms, Dunford, Kanter, Kr\"assig, Landahl, Pratt, Santra, Southworth
  et~al.}}]{Peterson:XR-up}
\bibinfo{author}{\bibfnamefont{E.~R.} \bibnamefont{Peterson}},
  \bibinfo{author}{\bibfnamefont{C.}~\bibnamefont{Buth}},
  \bibinfo{author}{\bibfnamefont{D.~A.} \bibnamefont{Arms}},
  \bibinfo{author}{\bibfnamefont{R.~W.} \bibnamefont{Dunford}},
  \bibinfo{author}{\bibfnamefont{E.~P.} \bibnamefont{Kanter}},
  \bibinfo{author}{\bibfnamefont{B.}~\bibnamefont{Kr\"assig}},
  \bibinfo{author}{\bibfnamefont{E.~C.} \bibnamefont{Landahl}},
  \bibinfo{author}{\bibfnamefont{S.~T.} \bibnamefont{Pratt}},
  \bibinfo{author}{\bibfnamefont{R.}~\bibnamefont{Santra}},
  \bibinfo{author}{\bibfnamefont{S.~H.} \bibnamefont{Southworth}},
  \bibnamefont{et~al.}, \bibinfo{journal}{submitted}  (\bibinfo{year}{2008}).

\bibitem[{\citenamefont{Szabo and Ostlund}(1989)}]{Szabo:MQC-89}
\bibinfo{author}{\bibfnamefont{A.}~\bibnamefont{Szabo}} \bibnamefont{and}
  \bibinfo{author}{\bibfnamefont{N.~S.} \bibnamefont{Ostlund}},
  \emph{\bibinfo{title}{Modern quantum chemistry: Introduction to advanced
  electronic structure theory}} (\bibinfo{publisher}{McGraw-Hill},
  \bibinfo{address}{New York}, \bibinfo{year}{1989}), \bibinfo{edition}{{1st,
  revised}} ed., ISBN \bibinfo{isbn}{0-486-69186-1}.

\bibitem[{\citenamefont{Kroto}(1975)}]{Kroto:MR-75}
\bibinfo{author}{\bibfnamefont{H.~W.} \bibnamefont{Kroto}},
  \emph{\bibinfo{title}{Molecular rotation spectra}} (\bibinfo{publisher}{John
  Wiley~\& Sons}, \bibinfo{address}{London}, \bibinfo{year}{1975}), ISBN
  \bibinfo{isbn}{0-471-50853-5}.

\bibitem[{\citenamefont{Rose}(1957)}]{Rose:ET-57}
\bibinfo{author}{\bibfnamefont{M.~E.} \bibnamefont{Rose}},
  \emph{\bibinfo{title}{Elementary theory of angular momentum}}, Structure of
  matter (\bibinfo{publisher}{John Wiley~\& Sons}, \bibinfo{address}{New York},
  \bibinfo{year}{1957}), ISBN \bibinfo{isbn}{0-486-68480-6}.

\bibitem[{\citenamefont{Zare}(1988)}]{Zare:AM-88}
\bibinfo{author}{\bibfnamefont{R.~N.} \bibnamefont{Zare}},
  \emph{\bibinfo{title}{Angular momentum}} (\bibinfo{publisher}{John Wiley~\&
  Sons}, \bibinfo{address}{New York}, \bibinfo{year}{1988}), ISBN
  \bibinfo{isbn}{0-471-85892-7}.

\bibitem[{\citenamefont{Reichl}(2004)}]{Reichl:SP-04}
\bibinfo{author}{\bibfnamefont{L.~E.} \bibnamefont{Reichl}},
  \emph{\bibinfo{title}{A modern course in statistical physics}}
  (\bibinfo{publisher}{Wiley-VCH}, \bibinfo{address}{Weinheim},
  \bibinfo{year}{2004}), \bibinfo{edition}{2nd} ed., ISBN
  \bibinfo{isbn}{0-471-59520-9}.

\bibitem[{\citenamefont{Blum}(1996)}]{Blum:DM-96}
\bibinfo{author}{\bibfnamefont{K.}~\bibnamefont{Blum}},
  \emph{\bibinfo{title}{Density matrix theory and applications}}, Physics of
  atoms and molecules (\bibinfo{publisher}{Plenum Press}, \bibinfo{address}{New
  York}, \bibinfo{year}{1996}), \bibinfo{edition}{2nd} ed., ISBN
  \bibinfo{isbn}{0-306-45341-X}.

\bibitem[{\citenamefont{Townes and Schawlow}(1955)}]{Townes:MS-55}
\bibinfo{author}{\bibfnamefont{C.~H.} \bibnamefont{Townes}} \bibnamefont{and}
  \bibinfo{author}{\bibfnamefont{A.~L.} \bibnamefont{Schawlow}},
  \emph{\bibinfo{title}{Microwave spectroscopy}}
  (\bibinfo{publisher}{McGraw-Hill}, \bibinfo{address}{New York},
  \bibinfo{year}{1955}), ISBN \bibinfo{isbn}{0-486-61798-X}.

\bibitem[{\citenamefont{Craig and Thirunamachandran}(1984)}]{Craig:MQ-84}
\bibinfo{author}{\bibfnamefont{D.~P.} \bibnamefont{Craig}} \bibnamefont{and}
  \bibinfo{author}{\bibfnamefont{T.}~\bibnamefont{Thirunamachandran}},
  \emph{\bibinfo{title}{Molecular quantum electrodynamics}}
  (\bibinfo{publisher}{Academic Press}, \bibinfo{address}{London},
  \bibinfo{year}{1984}), ISBN \bibinfo{isbn}{0-486-40214-2}.

\bibitem[{\citenamefont{Bishop}(1990)}]{Bishop:MV-90}
\bibinfo{author}{\bibfnamefont{D.~M.} \bibnamefont{Bishop}},
  \bibinfo{journal}{Rev. Mod. Phys.} \textbf{\bibinfo{volume}{62}},
  \bibinfo{pages}{343} (\bibinfo{year}{1990}).

\bibitem[{\citenamefont{Merzbacher}(1998)}]{Merzbacher:QM-98}
\bibinfo{author}{\bibfnamefont{E.}~\bibnamefont{Merzbacher}},
  \emph{\bibinfo{title}{Quantum mechanics}} (\bibinfo{publisher}{John Wiley~\&
  Sons}, \bibinfo{address}{New York}, \bibinfo{year}{1998}),
  \bibinfo{edition}{3rd} ed., ISBN \bibinfo{isbn}{0-471-88702-1}.

\bibitem[{\citenamefont{Goodbody}(1982)}]{Goodbody:CT-82}
\bibinfo{author}{\bibfnamefont{A.~M.} \bibnamefont{Goodbody}},
  \emph{\bibinfo{title}{Cartesian tensors: with applications to mechanics,
  fluid mechanics and elasticity}} (\bibinfo{publisher}{Ellis Horwood Limited},
  \bibinfo{address}{Chichester}, \bibinfo{year}{1982}), ISBN
  \bibinfo{isbn}{0-85312-377-2}.

\bibitem[{\citenamefont{Sobelman}(1979)}]{Sobelman:AS-79}
\bibinfo{author}{\bibfnamefont{I.~I.} \bibnamefont{Sobelman}},
  \emph{\bibinfo{title}{Atomic spectra and radiative transitions}},
  vol.~\bibinfo{volume}{1} of \emph{\bibinfo{series}{Springer series in
  chemical physics}} (\bibinfo{publisher}{Springer}, \bibinfo{address}{Berlin},
  \bibinfo{year}{1979}), ISBN \bibinfo{isbn}{3-540-09082-7}.

\bibitem[{foo({\natexlab{a}})}]{footnote1}
\bibinfo{note}{As $\mat\alpha^{\mathrm S}(\omega\I{L})$ is symmetric, there are
  no rank one contributions~\cite{Rose:ET-57,Sobelman:AS-79,Zare:AM-88}.}

\bibitem[{foo({\natexlab{b}})}]{footnote2}
\bibinfo{note}{The time-evolution operator is not given by the usual
  exponential form due to the time dependence of the interaction with the laser
  and the x~rays~\cite{Merzbacher:QM-98}.}

\bibitem[{\citenamefont{Golub and Ortega}(1992)}]{Golub:SC-92}
\bibinfo{author}{\bibfnamefont{G.~H.} \bibnamefont{Golub}} \bibnamefont{and}
  \bibinfo{author}{\bibfnamefont{J.~M.} \bibnamefont{Ortega}},
  \emph{\bibinfo{title}{Scientific computing and differential equations: an
  introduction to numerical methods}} (\bibinfo{publisher}{Academic Press},
  \bibinfo{address}{New York}, \bibinfo{year}{1992}), ISBN
  \bibinfo{isbn}{0-12-289255-0}.

\bibitem[{\citenamefont{Meystre and Sargent~III}(1991)}]{Meystre:QO-91}
\bibinfo{author}{\bibfnamefont{P.}~\bibnamefont{Meystre}} \bibnamefont{and}
  \bibinfo{author}{\bibfnamefont{M.}~\bibnamefont{Sargent~III}},
  \emph{\bibinfo{title}{Elements of quantum optics}}
  (\bibinfo{publisher}{Springer}, \bibinfo{address}{Berlin},
  \bibinfo{year}{1991}), \bibinfo{edition}{2nd} ed., ISBN
  \bibinfo{isbn}{3-540-54190-X}.

\bibitem[{\citenamefont{Cohen-Tannoudji
  et~al.}(1977)\citenamefont{Cohen-Tannoudji, Diu, and Lalo\"e}}]{Cohen:QM-77}
\bibinfo{author}{\bibfnamefont{C.}~\bibnamefont{Cohen-Tannoudji}},
  \bibinfo{author}{\bibfnamefont{B.}~\bibnamefont{Diu}}, \bibnamefont{and}
  \bibinfo{author}{\bibfnamefont{F.}~\bibnamefont{Lalo\"e}},
  \emph{\bibinfo{title}{Quantum mechanics}} (\bibinfo{publisher}{John Wiley \&
  Sons}, \bibinfo{address}{New York}, \bibinfo{year}{1977}), ISBN
  \bibinfo{isbn}{0-471-16432-1}.

\bibitem[{foo({\natexlab{c}})}]{footnote5}
\bibinfo{note}{The complex conjugation of~$\vec d^{\>\prime}_{ii'}$ in the
  Hermitian scalar product in Eq.~(\ref{eq:H_X_matel}) prevents that $\vec
  d^{\>\prime}_{ii'}$ is actually complex conjugated. This is necessary for the
  scalar product of the vectors to reproduce the matrix element between the
  direct product states~(\ref{eq:dirprod}) and the direct product of
  operators~$\hat{\vec d} \cdot \vec e\I{X,M}$ in
  Eq.~(\ref{eq:xrayint})~\cite{Cohen:QM-77}.}

\bibitem[{foo({\natexlab{d}})}]{footnote3}
\bibinfo{note}{Note that we define the cross correlation for~$t-\tau$ instead
  of~$t+\tau$ as is common in mathematics. According to our definition, a
  positive (negative)~$\tau$ implies that the x-ray pulse maximum comes later
  (earlier) than the laser pulse maximum.}

\bibitem[{foo({\natexlab{e}})}]{footnote4}
\bibinfo{note}{The orbital plots of~Br$_2$ were produced by the
  \textsc{molden}~\cite{Schaftenaar:MO-00} program based on a computation with
  the Hartree-Fock module of \textsc{dalton}~\cite{dalton:pgm-05} using the
  cc-pVDZ basis set~\cite{Wilson:GBS-99,basislib:02-02-06}.}

\bibitem[{\citenamefont{Schaftenaar and Noordik}(2000)}]{Schaftenaar:MO-00}
\bibinfo{author}{\bibfnamefont{G.}~\bibnamefont{Schaftenaar}} \bibnamefont{and}
  \bibinfo{author}{\bibfnamefont{J.~H.} \bibnamefont{Noordik}},
  \bibinfo{journal}{J. Comput.-Aided Mol. Design}
  \textbf{\bibinfo{volume}{14}}, \bibinfo{pages}{123} (\bibinfo{year}{2000}).

\bibitem[{\citenamefont{Filipponi and D'~Angelo}(1998)}]{Filipponi:AD-98}
\bibinfo{author}{\bibfnamefont{A.}~\bibnamefont{Filipponi}} \bibnamefont{and}
  \bibinfo{author}{\bibfnamefont{P.}~\bibnamefont{D'~Angelo}},
  \bibinfo{journal}{J. Chem. Phys.} \textbf{\bibinfo{volume}{109}},
  \bibinfo{pages}{5356} (\bibinfo{year}{1998}).

\bibitem[{\citenamefont{Bishop}(1973)}]{Bishop:GT-73}
\bibinfo{author}{\bibfnamefont{D.~M.} \bibnamefont{Bishop}},
  \emph{\bibinfo{title}{Group theory and chemistry}}
  (\bibinfo{publisher}{Clarendon Press}, \bibinfo{address}{Oxford},
  \bibinfo{year}{1973}), ISBN \bibinfo{isbn}{0-486-67355-3}.

\bibitem[{\citenamefont{Atkins and Friedman}(2004)}]{Atkins:MQM-04}
\bibinfo{author}{\bibfnamefont{P.~W.} \bibnamefont{Atkins}} \bibnamefont{and}
  \bibinfo{author}{\bibfnamefont{R.~S.} \bibnamefont{Friedman}},
  \emph{\bibinfo{title}{Molecular Quantum Mechanics}}
  (\bibinfo{publisher}{Oxford University Press}, \bibinfo{address}{Oxford},
  \bibinfo{year}{2004}), \bibinfo{edition}{forth} ed., ISBN
  \bibinfo{isbn}{0-19-927498-3}.

\bibitem[{\citenamefont{Archibong and Thakkar}(1993)}]{Archibong:SP-93}
\bibinfo{author}{\bibfnamefont{E.~F.} \bibnamefont{Archibong}}
  \bibnamefont{and} \bibinfo{author}{\bibfnamefont{A.~J.}
  \bibnamefont{Thakkar}}, \bibinfo{journal}{Chem. Phys. Lett.}
  \textbf{\bibinfo{volume}{201}}, \bibinfo{pages}{485} (\bibinfo{year}{1993}).

\bibitem[{\citenamefont{Thomas et~al.}(2002)\citenamefont{Thomas, Miron,
  Wiesner, Morin, Carroll, and S\ae{}thre}}]{Thomas:ANL-02}
\bibinfo{author}{\bibfnamefont{T.~D.} \bibnamefont{Thomas}},
  \bibinfo{author}{\bibfnamefont{C.}~\bibnamefont{Miron}},
  \bibinfo{author}{\bibfnamefont{K.}~\bibnamefont{Wiesner}},
  \bibinfo{author}{\bibfnamefont{P.}~\bibnamefont{Morin}},
  \bibinfo{author}{\bibfnamefont{T.~X.} \bibnamefont{Carroll}},
  \bibnamefont{and} \bibinfo{author}{\bibfnamefont{L.~J.}
  \bibnamefont{S\ae{}thre}}, \bibinfo{journal}{Phys. Rev. Lett.}
  \textbf{\bibinfo{volume}{89}}, \bibinfo{pages}{223001}
  (\bibinfo{year}{2002}).

\bibitem[{\citenamefont{Buth and
  Santra}(2007{\natexlab{b}})}]{fella:pgm-V1.2.0}
\bibinfo{author}{\bibfnamefont{C.}~\bibnamefont{Buth}} \bibnamefont{and}
  \bibinfo{author}{\bibfnamefont{R.}~\bibnamefont{Santra}},
  \emph{\bibinfo{title}{\textsc{fella} -- the free electron laser atomic,
  molecular, and optical physics program package}},
  \bibinfo{organization}{Argonne National Laboratory},
  \bibinfo{address}{Argonne, Illinois, USA}
  (\bibinfo{year}{2007}{\natexlab{b}}), \bibinfo{note}{version 1.2.0, with
  contributions by Mark Baertschy, Kevin Christ, Hans-Dieter Meyer, and Thomas
  Sommerfeld}.

\bibitem[{\citenamefont{Mills et~al.}(1988)\citenamefont{Mills, Cvitas, Homann,
  Kallay, and Kuchitsu}}]{Mills:QU-88}
\bibinfo{author}{\bibfnamefont{I.}~\bibnamefont{Mills}},
  \bibinfo{author}{\bibfnamefont{T.}~\bibnamefont{Cvitas}},
  \bibinfo{author}{\bibfnamefont{K.}~\bibnamefont{Homann}},
  \bibinfo{author}{\bibfnamefont{N.}~\bibnamefont{Kallay}}, \bibnamefont{and}
  \bibinfo{author}{\bibfnamefont{K.}~\bibnamefont{Kuchitsu}},
  \emph{\bibinfo{title}{Quantities, units and symbols in physical chemistry}}
  (\bibinfo{publisher}{Blackwell Scientific Publications},
  \bibinfo{address}{Oxford}, \bibinfo{year}{1988}), \bibinfo{edition}{2nd} ed.,
  ISBN \bibinfo{isbn}{0-632-03583-8}.

\bibitem[{\citenamefont{Rosman and Taylor}(1998)}]{Rosman:IC-98}
\bibinfo{author}{\bibfnamefont{K.~J.~R.} \bibnamefont{Rosman}}
  \bibnamefont{and} \bibinfo{author}{\bibfnamefont{P.}~\bibnamefont{Taylor}},
  \bibinfo{journal}{Pure Appl. Chem.} \textbf{\bibinfo{volume}{70}},
  \bibinfo{pages}{217} (\bibinfo{year}{1998}).

\bibitem[{\citenamefont{Christiansen et~al.}(1998)\citenamefont{Christiansen,
  Halkier, Koch, J\o{}rgensen, and Helgaker}}]{Christiansen:ID-98}
\bibinfo{author}{\bibfnamefont{O.}~\bibnamefont{Christiansen}},
  \bibinfo{author}{\bibfnamefont{A.}~\bibnamefont{Halkier}},
  \bibinfo{author}{\bibfnamefont{H.}~\bibnamefont{Koch}},
  \bibinfo{author}{\bibfnamefont{P.}~\bibnamefont{J\o{}rgensen}},
  \bibnamefont{and} \bibinfo{author}{\bibfnamefont{T.}~\bibnamefont{Helgaker}},
  \bibinfo{journal}{J. Chem. Phys.} \textbf{\bibinfo{volume}{108}},
  \bibinfo{pages}{2801} (\bibinfo{year}{1998}).

\bibitem[{dal()}]{dalton:pgm-05}
\bibinfo{note}{\textsc{dalton}, a molecular electronic structure program,
  Release~2.0 (2005), see
  \href{http://www.kjemi.uio.no/software/dalton/dalton.html}%
  {www.kjemi.uio.no/software/dalton/dalton.html}}.

\bibitem[{\citenamefont{Wilson et~al.}(1999)\citenamefont{Wilson, Woon,
  Peterson, and {Dunning, Jr.}}}]{Wilson:GBS-99}
\bibinfo{author}{\bibfnamefont{A.~K.} \bibnamefont{Wilson}},
  \bibinfo{author}{\bibfnamefont{D.~E.} \bibnamefont{Woon}},
  \bibinfo{author}{\bibfnamefont{K.~A.} \bibnamefont{Peterson}},
  \bibnamefont{and} \bibinfo{author}{\bibfnamefont{T.~H.}
  \bibnamefont{{Dunning, Jr.}}}, \bibinfo{journal}{J. Chem. Phys.}
  \textbf{\bibinfo{volume}{110}}, \bibinfo{pages}{7667} (\bibinfo{year}{1999}).

\bibitem[{bas()}]{basislib:02-02-06}
\bibinfo{note}{Basis sets were obtained from the \emph{Extensible Computational
  Chemistry Environment Basis Set Database}, Version 02/02/06, as developed and
  distributed by the Molecular Science Computing Facility, Environmental and
  Molecular Sciences Laboratory which is part of the Pacific Northwest
  Laboratory, P.O.~Box~999, Richland, Washington~99352, USA, and funded by the
  U.S. Department of Energy. The Pacific Northwest Laboratory is a
  multi-program laboratory operated by Battelle Memorial Institute for the
  U.S.~Department of Energy under contract~DE-AC06-76RLO~1830. Contact Karen
  Schuchardt for further information.}

\bibitem[{\citenamefont{Ramakrishna and Seideman}(2005)}]{Ramakrishna:IL-05}
\bibinfo{author}{\bibfnamefont{S.}~\bibnamefont{Ramakrishna}} \bibnamefont{and}
  \bibinfo{author}{\bibfnamefont{T.}~\bibnamefont{Seideman}},
  \bibinfo{journal}{Phys. Rev. Lett.} \textbf{\bibinfo{volume}{95}},
  \bibinfo{pages}{113001} (\bibinfo{year}{2005}).

\bibitem[{\citenamefont{St\"ohr}(1996)}]{Stohr:NE-96}
\bibinfo{author}{\bibfnamefont{J.}~\bibnamefont{St\"ohr}},
  \emph{\bibinfo{title}{NEXAFS Spectroscopy}}, vol.~\bibinfo{volume}{25} of
  \emph{\bibinfo{series}{Springer series in surface sciences}}
  (\bibinfo{publisher}{Springer}, \bibinfo{address}{New York},
  \bibinfo{year}{1996}), ISBN \bibinfo{isbn}{0-387-54422-4}.

\bibitem[{\citenamefont{Adachi et~al.}(2005)\citenamefont{Adachi, Kosugi, and
  Yagishita}}]{Adachi:SR-05}
\bibinfo{author}{\bibfnamefont{J.-i.} \bibnamefont{Adachi}},
  \bibinfo{author}{\bibfnamefont{N.}~\bibnamefont{Kosugi}}, \bibnamefont{and}
  \bibinfo{author}{\bibfnamefont{A.}~\bibnamefont{Yagishita}},
  \bibinfo{journal}{J. Phys. B} \textbf{\bibinfo{volume}{38}},
  \bibinfo{pages}{R127} (\bibinfo{year}{2005}).
\end{thebibliography}
\end{document}